\documentclass[pra,a4paper,twocolumn]{revtex4}
\usepackage[latin1]{inputenc}
\usepackage{amsmath}
\usepackage{amsfonts}
\usepackage{amssymb}
\usepackage{graphicx}
\newlength{\figwidtha} \newlength{\figwidthb}   \newcommand{\lc}{\left[} \newcommand{\rc}{\right]}   \newcommand{\ex}[1]{\bigl< #1 \bigr>}               

\date{\today}
\begin{document}

\title{Simulations and Experiments on Polarisation Squeezing in Optical Fibre} 

\author{Joel F. Corney$^1$, Joel Heersink$^2$, Ruifang Dong$^2$, Vincent Josse$^2$, Peter D. Drummond$^1$, Gerd Leuchs$^2$ and Ulrik L. Andersen$^{2,3}$}

\affiliation{$^1$ARC Centre of Excellence for Quantum-Atom Optics, School of Physical
Sciences, The University of Queensland, Brisbane, QLD 4072, Australia}
\email{corney@physics.uq.edu.au}

\affiliation{$^2$Institut f\"ur Optik, Information und Photonik, Max-Planck Forschungsgruppe,
Universit\"at Erlangen-N\"urnberg, G\"unther-Scharowsky-Strasse 1, 91058 Erlangen, Germany}
\affiliation{$^3$Department of Physics, Technical University of Denmark, Building 309, 2800 Kgs.\ Lyngby, Denmark}

\begin{abstract}
We investigate polarisation squeezing of ultrashort pulses in optical fibre, over a wide range of input energies and fibre lengths.  Comparisons are made between experimental data and quantum dynamical simulations, to find good quantitative agreement.  The numerical calculations, performed using both truncated Wigner and exact $+P$ phase-space methods, include nonlinear and stochastic Raman effects, through coupling to phonons variables.  The simulations reveal that excess phase noise, such as from depolarising GAWBS, affects squeezing at low input energies, while Raman effects cause a marked deterioration of squeezing at higher energies and longer fibre lengths. The optimum fibre length for maximum squeezing is also calculated.\end{abstract}
\maketitle

\section{Introduction}


The search for efficient means of quantum squeezing, in which quantum fluctuations in one observable are reduced below the standard quantum limit, at the expense of increased fluctuations in the conjugate,  has been at the heart of modern developments in quantum optics\cite{Drummond:2004}.  As well as for the fundamental interest of highly nonclassical light,  optical squeezing is of interest  for quantum information applications. Possible uses include:  generating entanglement for quantum communication\cite{braunstein03.book}, making measurements below the standard quantum limit\cite{giovannetti04.sci}, and for precise engineering of the quantum states of matter\cite{hald00.jomo}.

The use of optical fibre for quantum squeezing has considerable technological advantages, such as generating squeezing directly at the communications wavelength and use of existing transmission technology.  There is, however, a significant disadvantage in the excess phase noise that arises from acoustic waves, molecular vibrations, and defects in the amorphous silica.  

Here we present an in-depth numerical and experimental study of polarisation squeezing in a single-pass scheme that successfully reduces the impact of this excess phase noise. The numerical simulations represent a quantitative, experimentally testable solution of quantum many-body dynamics.

The first proposals for the generation of squeezed light using the $\chi^{(3)}$ nonlinearity date back to 1979, with schemes involving a nonlinear Kerr interferometer~\cite{ritze79.oc} or degenerate four-wave mixing~\cite{yuen79.ol}. The first experimental demonstration used four-wave mixing in atomic samples~\cite{slusher85.prl}. The Kerr effect in optical fibers was also proposed as a mechanism for squeezing light~\cite{levenson85.ol,levenson85.pra,kitagawa86.pra}. Squeezing using fibres was first successfully implemented using a continuous wave laser, and was observed by a phase shifting cavity~\cite{shelby86.prl}. 

However, early experiments\cite{levenson85.ol,levenson85.pra,shelby86.prl} were severely limited by the phase noise intrinsic to optical fibre.  Such noise occurs in the form of thermally excited refractive index fluctuations in the fiber~\cite{shelby85.prl, perlmutter90.prb}, and arises from Guided Acoustic Wave Brillouin Scattering (GAWBS) and $1/f$ noise.  
A substantial theoretical breakthrough was the recognition that short pulses - ideally  in the form of solitons - could lead to much higher peak powers, thus allowing the generation of nonclassical light in with fiber lengths short enough so that thermally induced phase noise was not an issue.  Such short pulses required a true multi-mode theoretical approach\cite{Carter:1987p1841}, which led to the first predictions of pulsed squeezing, and to an understanding of the scaling laws involved \cite{Shelby1990a}.

These predictions were confirmed in a landmark experiment by Rosenbluh and Shelby~\cite{rosenbluh91.prl} , which used intense, sub-picosecond laser pulses to eliminate much of the phase noise, and a simpler interferometric setup~\cite{kitagawa86.pra} in a balanced configuration.  All fiber squeezers since have exploited ultrashort pulses.
Observation schemes implemented with standard fibers include: i) phase-shifting cavities~\cite{shelby86.prl}, ii) spectral filtering~\cite{friberg96.prl, spaelter98.oe, koenig98.jomo, nishizawa00.jjap, takeoka01.ol}, iii) balanced interferometers~\cite{rosenbluh91.prl, bergman91.ol, bergman93.ol, bergman94.ol, yu01.ol}, iv) asymmetric interferometers~\cite{schmitt98.prl, krylov98.ol, fiorentino2001a, heersink03.pra, meissner04.job} and v) a two-pulse, single-pass method generating squeezed vacuum~\cite{margalit98.oe, nishizawa02.jjap} or polarization squeezing~\cite{heersink05.ol,Dong:2008p116}. 


 Squeezing the polarization variables of light is a promising alternative~\cite{korolkova02.pra} to the squeezing in the amplitude quadrature or the photon number, which the vast majority of fiber squeezing experiments until now have implemented. That the quantum polarization variables could also exhibit noise reduction was first suggested by Chirkin~\textit{et al.} in 1993~\cite{chirkin93.qe}. This proposal for polarization squeezing is similar to earlier remarks by Walls and Zoller concerning atomic spin squeezing~\cite{walls81.prl} due to the systems' mathematical similarities. The first experiment to exploit the quantum properties of polarization was performed by Grangier~\textit{et al.} in 1987 in a squeezed-light-enhanced polarization interferometer~\cite{grangier87.prl}. The first explicit demonstration was achieved by S{\o}rensen~\textit{et~al.} in the context of quantum memory~\cite{sorensen98.prl}. Such a promising application sparked intensified interest, resulting in a number of theoretical investigations, e.g.~\cite{alodjants98.apb, korolkova02.pra, luis02.pra}. These in turn spawned a plethora of experiments in a variety of different systems: optical parametric oscillators~\cite{bowen02a.prl, bowen02b.prl, schnabel03.pra}, optical fibers~\cite{heersink03.pra, heersink05.ol,Dong:2008p116} and cold atomic samples~\cite{josse03.prl}. 

In this paper we present a detailed experimental  and theoretical investigation of the single-pass method for creating polarization squeezing, building upon our previous work~\cite{heersink05.ol, Corney:2006p023606}. This efficient and novel squeezing source has a number of advantages compared with previous experiments producing bright squeezing. For example, this setup is capable of producing squeezing over a wide range of powers, in contrast to asymmetric Sagnac loop schemes. There is thus a certain similarity to experiments using a Mach Zehnder interferometer as a flexible asymmetric Sagnac loop~\cite{fiorentino2001a}. The interference of a strong squeezed and a weak `coherent' beam in asymmetric loops however gives rise to a degradation in squeezing due to the dissimilarity of the pulses as well as losses from the asymmetric beam splitter.

In the single-pass scheme, the destructive effect arising from interfering dissimilar pulses (in power, temporal and spectral shape) is avoided by interfering two strong Kerr-squeezed pulses that co-propagate on orthogonal polarization axes.  For equal power they have been found to be virtually identical within measurement uncertainties in, e.g. spectrum, autocorrelation and squeezing. This scheme presents the potential to measure greater squeezing and provides a greater robustness against input power fluctuations. Formally this interference of equally squeezed pulses is reminiscent of earlier experiments producing vacuum squeezing, for example~\cite{rosenbluh91.prl, bergman91.ol, nishizawa02.jjap}. The advantage here is that no extra local oscillator is needed in the measurement of polarization squeezing. 

These novel experiments allow a careful experimental test of the multi-mode theory of optical squeezing.   Here we make use of the comprehensive model developed by Carter and Drummond\cite{Carter:1991p3757} that includes the electronic $\chi^{(3)}$ nonlinear responses of the material and nonresonant coupling to phonons in the silica.  The phonons provide a non-Markovian reservoir that generates additional, delayed nonlinearity, as well as spontaneous and thermal noise.  The coupling is based on the experimentally determined Raman gain $\alpha^{R}(\omega)$~\cite{Stolen1989a}.  

The simulation of pulse propagation entails the solution of time-domain dynamical evolution in a quantum field theory with large numbers of interacting particles.   We achieve this here primarily with a truncated Wigner technique\cite{Drummond:1993p279}, which provides an accurate simulation of the quantum dynamics for short propagation times and large photon number. The quantum effects enter via initial vacuum noise, which makes the technique ideally suited to squeezing calculations.  We compare simulation and experiments to find excellent agreement over a wide range of pulse energies and fibre lengths.  From the simulations, we can identify the particular noise sources that are the limiting factors at high and low input energy.

We begin in Sec.~\ref{Squeezing} with an introduction to polarisation squeezing by means of the Kerr effect, from a single-mode picture, before presenting the detailed model of pulse propagation in fibres in Sec.~\ref{Pulse}.  Sections \ref{Simulations} and \ref{Outputs} describe the numerical simulation methods used, while the experimental set-up is described in Sec.~\ref{Experiment}.  Sec.~\ref{Results} discusses the results of both the experiment and simulations.  The appendices contain further details of the theoretical description and numerical simulation.


\section{Squeezing}
\label{Squeezing}

\subsection{Kerr squeezing}

The generation of squeezed optical beams requires a nonlinear interaction to transform the statistics of the input, which is typically a coherent state. The first observation of quantum noise squeezing used four wave mixing~\cite{slusher85.prl} arising from the third order electric susceptibility $\chi^{(3)}$. Although material dispersion can place limits on the interaction length, this limitation can be circumvented by use of degenerate frequencies\cite{shelby86.pra}, as in the optical Kerr effect. Here the interaction has the effect of introducing an intensity dependence to the medium's refractive index, Eq.~(\ref{eq_n2}), which in turn induces an intensity-dependent phase shift in incident pulses. This effect dominates the nonlinearity in fibers made of fused silica, a material with an inversion symmetric molecular ordering. In a pure Kerr material the refractive index is an instantaneous function of the optical intensity and the refractive index $n$ is then given to second order by~\cite{sizmann99.pio}:
\begin{equation}
  n = n_0 +n_2 I \quad{} \text{with}
  \quad{} n_2 = \frac{3}{4}\ \frac{{\rm Re}\left(\chi^{(3)}_{xxxx}\right)}{n_{0}^2 \epsilon_{0} c},
\label{eq_n2}
\end{equation}
where the optical intensity is given by \mbox{$I = \frac{1}{2} n_0 \epsilon_0 c |E|^2$} and $\chi^{(3)}_{xxxx}$ is the third order susceptibility coefficient for the degenerate mode $x$. The instantaneity of fused silica's nonlinearity is true only to a first approximation. In reality, it is only the electronic contribution,  which typically comprises 85\% of the total nonlinearity~\cite{smolorz99.ol}, that is instantaneous on the scale of the 130~fs pulses used here. The time dependence of the remainder cannot be neglected and arises primarily from Raman scattering~\cite{stolen89.josab}. Nonetheless, the simplification of an instantaneous response can be useful in gaining physical insight into the Kerr squeezing mechanism. 

\begin{figure}[h]
\centering
\includegraphics[scale=0.6]{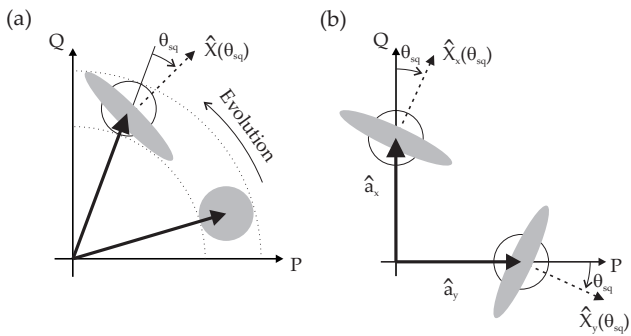}
\caption{\it (a) Representation in phase space of the evolution of a coherent beam (bottom right) under effect of the Kerr nonlinearity, which generates a quadrature (or Kerr) squeezed state (upper left). The arrow indicates the direction of state evolution with propagation. (b) Polarization squeezing generated by overlapping two orthogonally (i.e. $x$- and $y$-) polarized quadrature-squeezed states.} \label{fig_kerrsq}
\end{figure}

Figure~\ref{fig_kerrsq}(a) illustrates the effect of an instantaneous nonlinear refraction. Sending an ensemble of identical coherent states into a perfect Kerr medium causes a distortion of the initially symmetric phase-space distribution. One can explain this distortion by considering the input to consist of a superposition of photon number states, which the Kerr effect rotates relative to one another in phase space. The initially symmetric phase-space distribution characteristic of coherent states is thereby distorted into an ellipse or `squeezed' circle. Generally the squeezed state will be crescent shaped, however for the experimental conditions of high intensities and small nonlinearities our states never become significantly curved.

The resultant quantum state is quadrature squeezed, where the squeezed quadrature  $\hat{X}(\theta_{sq})$ is rotated by $\theta_{sq}$ relative to the amplitude quadrature or radial direction. The state's phase-space uncertainty distribution is altered such that the statistics in the amplitude quadrature remain constant in keeping with energy conservation. Thus the squeezed or noise-reduced optical quadrature cannot be detected directly in amplitude or intensity measurements. A detection scheme sensitive to the angle of the squeezed ellipse $\theta_{sq}$ is required. 

\subsection{Single-mode picture of polarisation squeezing}

\begin{figure}[h]
\begin{center}
\includegraphics[scale=0.7]{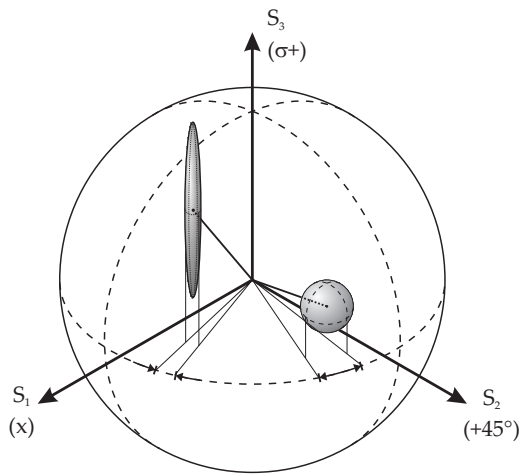}
\caption{\it Representation of the variances of a polarization squeezed (upper left) and a coherent state (lower right) on the Poincar\'e sphere.}\label{fig_poincaresphere}
\end{center}
\end{figure}

The characterization of quantum polarization states relies on the measurement of the quantum Stokes operators (see Ref.~\cite{korolkova02.pra} and references therein). These Hermitian operators are defined analogously to their classical counterparts as~\cite{jauch55.book}:
\begin{eqnarray}
\hat{S}_0&=&\hat{a}^{\dag}_{x}\hat{a}_{x}+\hat{a}^{\dag}_{y}\hat{a}_{y}, \quad
\hat{S}_1=\hat{a}^{\dag}_{x}\hat{a}_{x}-\hat{a}^{\dag}_{y}\hat{a}_{y},  \nonumber \\
\hat{S}_2&=&\hat{a}^{\dag}_{x}\hat{a}_{y}+\hat{a}^{\dag}_{y}\hat{a}_{x}, \quad
\hat{S}_3=i(\hat{a}^{\dag}_{y}\hat{a}_{x}-\hat{a}^{\dag}_{x}\hat{a}_{y}),
\label{eq_qstokes}
\end{eqnarray}
where $\hat{a}_x$ and $\hat{a}_y$ are two orthogonally polarized modes (with temporal,  position and mode dependence implicit). These operators obey the SU(2) Lie algebra and thus, within a factor of $\frac{\hbar}{2}$, coincide with the angular momentum operators. The commutators of these operators, following from the noncommutation of the photon operators, are given by:
\begin{equation}
\left[\hat{S}_0,\hat{S}_i \right] = 0 \quad \text{and} \quad  \left[\hat{S}_i,\hat{S}_j\right] =2i\epsilon_{ijk}\hat{S}_k,
\label{eq_qstoke_commutation}
\end{equation}
where $i,j,k = 1,2,3$ and where $\epsilon$ is the antisymmetric symbol. These commutation relations lead to Heisenberg inequalities and therefore to the presence of intrinsic quantum uncertainties in analog to those of the quadrature variables. However, the fundamental noise limit depends on the mean polarization state:
\begin{equation}
\Delta^2\hat{S}_i\Delta^2\hat{S}_j\geq \epsilon_{ijk} \left|\langle\hat{S}_k\rangle\right|^2,
\label{eq_stokeuncertainty}
\end{equation}
where the variance of $\hat{S}_i$ is given by \mbox{$\Delta^2\hat{S}_i=\langle\hat{S}_i^2\rangle-\langle\hat{S}_i\rangle^2$}.  This quantum picture of the polarization state of light cannot be represented as a point on the Poincar\'e sphere, but rather as a distribution in the space spanned by the Poincar\'e parameters, analogous to the phase-space representation of quantum optical states. This is visualized in Fig.~\ref{fig_poincaresphere},  which shows the variances, i.e. full-width at half-maximum of the marginal distributions, of a coherent and a polarization squeezed state.

Despite the fact that the Stokes uncertainty relations are state dependent, it is always possible to find pairs of maximally conjugate operators. This is equivalent to defining a Stokes basis in which only one parameter has a nonzero expectation value. This is justified insomuch that polarization transformations are unitary. Consider a polarization state described by \mbox{$\langle\hat{S}_i\rangle=\langle\hat{S}_j\rangle=0$} and \mbox{$\langle\hat{S}_k\rangle=\langle\hat{S}_0\rangle\neq0$} where $i,j,k$ represent orthogonal Stokes operators. The only nontrivial Heisenberg inequality then reads:
\begin{equation}
\Delta^2 \hat{S}_i\, \Delta^2 \hat{S}_j \geq |\langle\hat{S}_k\rangle|^2=|\langle\hat{S}_0\rangle|^2,
\end{equation}
which mirrors the quadrature uncertainty relation, and polarization squeezing can then be similarly defined:
\begin{equation}
\Delta^2 \hat{S}_i < |\langle\hat{S}_k\rangle| < \Delta^2 \hat{S}_j.
\label{eq_polsq}
\end{equation}
The definition of the conjugate operators $\hat{S}_i,\hat{S}_j$ is not unique and there exists an infinite set of conjugate operators $\hat{S}_\perp(\theta),\hat{S}_\perp(\theta+\frac{\pi}{2})$ that are perpendicular to the state's classical excitation $\hat{S}_k$, for which $\langle\hat{S}_\perp(\theta)\rangle =0$ for all $\theta$. All these operator pairs exist in the $\hat{S}_i-\hat{S}_j$ `dark plane,' i.e. the plane of zero mean intensity. A general dark plane operator is described by:
\begin{equation}
\hat{S}_{\perp}(\theta) \,=\, \cos(\theta)\hat{S}_i + \sin(\theta)\hat{S}_j,
\label{rotation}
\end{equation}
where $\theta$ is an angle in this plane defined relative to $\hat{S}_i$. Polarization squeezing is then generally given by:
\begin{equation}
\Delta^2 \hat{S}_\perp(\theta_{sq}) < |\langle\hat{S}_0\rangle| < \Delta^2 \hat{S}_\perp(\theta_{sq}+\frac{\pi}{2}),
\label{eq_stokes_uncertainty}
\end{equation}
where $\hat{S}_\perp(\theta_{sq})$ is the maximally squeezed parameter and $\hat{S}_\perp(\theta_{sq}+\frac{\pi}{2})$ the antisqueezed parameter. 

Consider, for example, the specific case of a $+\hat{S}_3$ or $\sigma_+$-polarized beam as in the experiments presented here. Let this beam be composed of the two independent modes $\hat{a}_x,\hat{a}_y$ with a relative $\frac{\pi}{2}$ phase shift between their mean values. This is depicted in Fig.~\ref{fig_kerrsq}(b) and described by $\langle\hat{a}_y\rangle=i\langle\hat{a}_x\rangle=i\alpha/\sqrt{2}$ and $\alpha \in\mathbb{R}$. The beam is then circularly polarized with $\hat{a}_{\sigma_+}$ as the mean field and $\hat{a}_{\sigma_-}$ is the orthogonal vacuum mode:
\begin{eqnarray}
\hat{a}_{\sigma_+} &=& \frac{-1}{\sqrt{2}} \left( \hat{a}_x - i\hat{a}_y \right) \quad \text{with} \quad \langle\hat{a}_{\sigma_+} \rangle = - \alpha, \nonumber \\
\hat{a}_{\sigma_-} &=& \frac{1}{\sqrt{2}} \left( \hat{a}_x + i\hat{a}_y \right) \quad \text{with} \quad \langle\hat{a}_{\sigma_+} \rangle = 0.
\end{eqnarray}

The Stokes operators in the plane spanned by $\hat{S}_1-\hat{S}_2$ correspond to the quadrature operators of the dark $\hat{a}_{\sigma_-}$-polarization mode. Assuming $|\langle\delta\hat{a}\rangle|\ll|\alpha|$ and considering only the noise terms, we find:
\begin{eqnarray}
\delta \hat{S}_\perp(\theta) &=& \alpha \left(\delta\hat{a}_{\sigma_-}e^{-i\theta} + \delta\hat{a}^\dagger_{\sigma_-}e^{i\theta} \right) \nonumber \\ 
&=& \alpha\delta\hat{X}_{\sigma_-}(\theta) \nonumber \\ 
&=& \alpha\left(\delta\hat{X}_x(\theta) + \delta\hat{X}_y(\theta-\frac{\pi}{2})\right), 
\label{eq_polsq_darkmode}
\end{eqnarray}
where the Stokes operator definitions of Eq.~(\ref{eq_qstokes}) have been used in a linearized form. The sum signal, a measure of the total intensity, is given by:
\begin{equation}
\delta \hat{S}_0 = \alpha\left(\delta\hat{a}_{\sigma_+} + \delta\hat{a}^\dagger_{\sigma_+} \right)= \alpha\delta\hat{X}_{\sigma_+},
\label{eq_polsq_brightmode}
\end{equation}
and thus exhibits no dependence on the dark mode. This considering of the physical interpretation of polarization squeezing shows that polarization squeezing is equivalent to vacuum squeezing in the orthogonal polarization mode:
\begin{equation}
\Delta^{2} \hat{S}_\perp(\theta) \;<\; |\langle\alpha\rangle|^2 \quad{} \Leftrightarrow \quad{} \Delta^2 \hat{X}_{\sigma_-}(\theta) \;<\; 1.
\label{eq_sq_equivalence}
\end{equation}

Whilst a particular case is considered here, a straightforward generalization to all other polarization bases is readily made as polarization transformations are unitary rotations in SU(2) space.

In dark-plane Stokes measurements, the beam's intensity is  divided equally between two photodetectors. Such measurements are then identical to balanced homodyne detection: the classical excitation is a local oscillator for the orthogonally polarized dark mode. The phase between these modes is varied by rotating the Stokes measurement through the dark plane, allowing full characterization of the noise properties of the dark, $y$-polarized mode. This is a unique feature of polarization measurements and has been used to great advantage in many experiments, for example~\cite{grangier87.prl, smithey93.prl, sorensen98.prl, julsgaard04.nat, bowen03.job, josse04.job, heersink05.ol, Heersink:2006p253601}. This has also allowed the first characterization of a bright Kerr squeezed state as well as the reconstruction of the polarization variable Q function using polarization measurements~\cite{Marquardt:2007p220401}.

To show how an $\hat S_3$ polarised state is squeezed by the Kerr effect, we consider the essential Kerr Hamiltonian:
\begin{equation}
\hat{H} = (\hat a^\dagger_x\hat a_x)^2 + (\hat a^\dagger_y\hat a_y)^2, 
\end{equation}
which in terms of the Stokes operators, can be expressed as
\begin{equation}
\hat{H} = \frac{1}{2} \left\{ \hat S_0^2 + \hat S_1^2 \right \}.
\end{equation}
The first term is a constant of the motion, since $\hat S_0$ gives the total number of photons, and has no effect on the dynamics.  The second term is a nonlinear precession around the $S_1$ axis: the rate of precession is proportional to $S_1$, which is a manifestation of the intensity-dependent refractive index of the Kerr effect. The nonlinear precession will distort an initially symmetric distribution centred in the $S_1$-$S_2$ plane (eg the $S_3$ circularly polarised state located at a pole of the sphere) into an ellipse.  As for ordinary quadrature squeezing, the nonlinear precession preserves the width in the $S_1$ direction,  and so the squeezing is not directly observable by a number-difference observation.

The advantage of the squeezed $S_3$ state, as opposed to squeezing of a linearly polarised  $S_2$  state, is that a simple rotation around the $S_3$ axis allows the squeezed (or antisqueezed) axis of the ellipse to be aligned to the $S_1$ axis and thus to be detected with a number-difference measurement.  Such a rotation is easily implemented experimentally with a polarisation rotator.

\section{Pulse propagation}
\label{Pulse}
\subsection{Multimode description}
We have so far described the polarisation squeezing as a single-mode Kerr effect.   However, this is accurate only for CW radiation, corresponding to a single momentum component.  Ultrashort pulses, on the other hand, correspond to a superposition of many plane waves and thus require a multimode description.  Such a description is crucial for an accurate treatment of dispersive and Raman effects.  For a continuum of momentum modes, we can express the superposition as
\begin{equation}
\widehat{\Psi}_{\sigma}(t,z)=\frac{1}{\sqrt{2\pi}}\int dk\,\widehat{a}_{\sigma}(t,k)e^{i(k-k_{0})z+i\omega_{0}t}, 
\end{equation}
where instead of annihilation and creation operators for each polarisation mode, we now have field operators $\widehat{\Psi}^\dagger_{\sigma}(t,z)$,  $\widehat{\Psi}_{\sigma}(t,z)$ for the envelopes of each of the polarisation modes $\sigma=(x,y)$.  The commutation relations of the fields are
\begin{equation}
\left[\widehat{\Psi}_{\sigma}(t,z),\widehat{\Psi}_{\sigma'}^{\dagger}(t,z')\right]=\delta(z-z')\delta_{\sigma\sigma'}, 
\end{equation}
and with this normalisation, the total number of $\sigma$-photons in the fibre is thus $\widehat{N}_\sigma (t)=\int_0^L  dz \widehat{\Psi}^\dagger_{\sigma}(t,z)\widehat{\Psi}_{\sigma}(t,z)$.

The general quantum model for a fibre with a single transverse mode is derived in \cite{Drummond:2001p139}.  The relevant aspects for the current system include the dispersive pulse propagation, the electronic polarisation response that gives the instantaneous $\chi^{(3)}$,  and the nonresonant coupling to phonons in the silica. 

\subsection{Electromagnetic Hamiltonian}
In terms of the field operators for the slowly varying envelope defined above, the normally ordered Hamiltonian for an electromagnetic pulse in a polarisation-preserving fibre under the rotating-wave approximation is:
\begin{eqnarray}
\widehat H_{\rm EM} &=&\hbar \sum_\sigma\int\int dz\; dz' \omega_\sigma(z-z') \widehat{\Psi}^\dagger_{\sigma}(t,z)\widehat{\Psi}_{\sigma}(t,z')  \nonumber \\
 && - \hbar \chi_E \sum_\sigma \int dz  \widehat{\Psi}^{\dagger2}_{\sigma}(t,z)\widehat{\Psi}^2_{\sigma}(t,z), 
\end{eqnarray}
where $ \omega_\sigma (z)$ is the Fourier transform of the dispersion relation:
\begin{eqnarray}
 \omega_\sigma(z) \equiv \frac{1}{2\pi} \int dk \omega_\sigma(k) e^{i(k-k_{0})z}, \end{eqnarray}
and $ \chi_E $ is the strength of the third-order polarisation response.  The birefringence of the polarisation response means that there are differences between the dispersion relations  $\omega_x$ and $\omega_y$.  The $\chi^{(3)}$ term is assumed to be independent of polarisation, and cross-Kerr effects are neglected, as the different group velocities of the pulses mean that the length of time that the pulses overlap in the fibre is negligible.  The fibre is assumed to be homogeneous, with both $\omega_\sigma(k)$ and $\chi_E$ independent of the distance $z$ down the fibre.

To simplify the description of the dispersive part, we Taylor expand $\omega(k)$ around $k=k_0$ up to second order, which introduces the group velocity $v\equiv d\omega/dk |_{k=k_0}$ and dispersion parameter $\omega'' \equiv d^2\omega/dk^2 |_{k=k_0}$.  Subtracting off the free evolution at the carrier frequency $\omega_0 = \omega_x(k_0)$, one obtains the simplified Hamiltonian:
\begin{eqnarray}
\widehat H_{\rm EM}' & = & \hbar \sum_\sigma\int dz \left\{  \frac{iv_\sigma}{2} (\nabla \widehat{\Psi}^\dagger_{\sigma}\widehat{\Psi}_{\sigma} - \widehat{\Psi}^\dagger_{\sigma}\nabla\widehat{\Psi}_{\sigma}) \right. \nonumber\\
&&+ \left. \frac{\omega''}{2}\nabla \widehat{\Psi}^\dagger_{\sigma}\nabla\widehat{\Psi}_{\sigma} 
 - \hbar \chi_E  \widehat{\Psi}^{\dagger2}_{\sigma}\widehat{\Psi}^2_{\sigma}\right\}.
\end{eqnarray}
Here we have not included the difference in phase velocity between the two polarisations, which just leads to a constant relative phase shift.

For the methods that we use in this paper, it is convenient to treat the quantum dynamics in the Heisenberg picture, with time-evolving field operators.   The equation of motion of the field annihilation operator that arises from the electromagnetic Hamiltonian is
\begin{eqnarray}
\frac{d}{dt} \widehat{\Psi}_{\sigma} &=& \frac{-i}{\hbar}\left[\widehat{\Psi}_{\sigma},\widehat H_{\rm EM}'  \right] \nonumber\\ 
& = & \left \{ - v_\sigma\nabla + \frac{i\omega''}{2}\nabla^2 + i\chi_E \widehat{\Psi}_{\sigma}\widehat{\Psi}^\dagger_{\sigma}\right\}\widehat{\Psi}_{\sigma}.
\end{eqnarray}

\subsection{Raman Hamiltonian}
As well as the interaction with electrons that produces the polarisation response, the radiation field also interacts with phonons in the silica.  The photons can excite both localised oscillations of the atoms around their equilibrium positions (Raman effect) as well as guided acoustic waves (GAWBs) along the waveguide.  The latter can be treated as a low-frequency component of the Raman spectrum, and produces random fluctuations in the refractive index.  However the effect of this is largely removed in this experiment through common-mode rejection, and any residual phase-noise can be accounted for by simple scaling laws (see section \ref{sec:phasenoise}).

The Raman interactions produce both excess phase noise and an additional nonlinearity.  The atomic oscillation is modelled as a set of harmonic oscillators at each point in the fibre, and is coupled to the radiation field by a simple dispersive interaction:

\begin{eqnarray}
\widehat H_{\rm R} &=&\hbar \sum_{\sigma,k} R_k \int dz \widehat{\Psi}^\dagger_{\sigma}(z)\widehat{\Psi}_{\sigma}(z) \left\{\widehat b_{\sigma k}(z) +  \widehat b^\dagger_{\sigma k}(z)  \right\}  \nonumber \\
 && + \hbar\sum_{\sigma,k}  \omega_k \widehat b^\dagger_{\sigma k}(z) \widehat b_{\sigma k}(z),
\end{eqnarray} 
where the phonon operators have the commutation relations   
 \begin{eqnarray}
\left[\hat{b}_{\sigma k}(z,t),\hat{b}_{\sigma' k'}^{\dagger}(z',t)\right] & = & \delta(z-z')\delta_{k,k'}\delta_{\sigma,\sigma'}
\end{eqnarray}

The spectral profile of this interaction $R(\omega)$ is well known from experimental measurements\cite{stolen89.josab} and is sampled here by oscillators of equal spectral spacing $\Delta\omega = \omega_{k+1}-\omega_k$, such that $\lim_{\Delta\omega \rightarrow 0} R_k/\sqrt{\Delta\omega} = R(\omega_k)$.  The finite spectral width of the Raman profile means that the Raman contribution to the nonlinearity is not instantaneous on the time-scale of the optical pulse, leading to such effects as the soliton frequency shift\cite{mitschke86.ol,gordon86.ol}.

With the Raman and electromagnetic Hamiltonians combined, one can derive complete Heisenberg operator equations of motion for the optical field operator and the phonon operators\cite{Drummond:2001p139}: \begin{eqnarray}
\left(\frac{\partial}{\partial t}+v\frac{\partial}{\partial z}\right)\hat{\Psi}(z,t) & = & \lc-i\sum_{k}R_{k}\left\{ \hat{b}_{k}+\hat{b}_{k}^{\dagger}\right\} \right.\nonumber\\
&&\left. +\frac{i\omega''}{2}\frac{\partial^{2}}{\partial z^{2}}
+i\chi^{E}\hat{\Psi}^{\dagger}\hat{\Psi}\rc\hat{\Psi}\nonumber\\
\frac{\partial}{\partial t}\hat{b}_{k}(z,t) & = & -i\omega_k\hat{b}_{k}-iR_{k}\hat{\Psi}^{\dagger}\hat{\Psi},\end{eqnarray}
 where we have suppressed the polarisation index, since the equations for each polarisation are independent.  
 
The initial state of the phonons is thermal, with 
\begin{equation}
\ex{\hat{b}_{k'}^{\dagger}(z',0)\hat{b}_{k}(z,0)}=n_{\textrm{{th}}}(\omega_k)\delta_{k,k'}\delta(z-z'),
\end{equation}
where $n_{\textrm{th}}(\omega)  = 1/\left[ \exp(\hbar\omega/kT)-1\right]$ is the Bose-Einstein distribution.

\section{Simulation Methods}
\label{Simulations}

\subsection{Phase-space methods}

Phase-space methods are a means of simulating the dynamics of multimode many-body quantum systems.  They are based on (quasi)probabilistic representations of the density matrix that are defined by means of coherent states.  Because they are based on coherent states, they are ideally suited to simulating quantum optical experiments, which in so many cases begin with the coherent output of a laser.  The two representations that give rise to practical numerical methods are the $+P$ \cite {Glauber:1963p2766,Sudarshan:1963p277,Chaturvedi:1977p187,Drummond:1980p2353} and Wigner
 \cite{Wigner:1932p749} distributions.  In both methods, the resultant description has the same structure as the mean-field, or classical, description, which is a form of nonlinear Schr\"odinger equation in the case of optical fibres.  However there are also additional quantum noise terms, which may appear in the initial conditions or in the dynamical equations.

The $+P$ method provides an \emph{exact} probabilistic description in which stochastic averages correspond to normally ordered correlations.  Because of this normal ordering, it is suited to intensity correlation measurements.  Quantum effects enter by stochastic terms that have the form of spontaneous scattering.  The $+P$ method has been applied to a variety of quantum-optical applications, including
superfluorescence\cite{Haake:1979p1740,Drummond:1982p3446}, parametric amplifiers\cite{Gardiner:2004} and optical fibres\cite{Carter:1987p1841,Carter:1991p3757}.  More recently, it has been applied to a variety of Bose-Einstein condensate (BEC) simulations\cite{Drummond:1999p2661,Drummond:2004p040405,Poulsen:2001p013616,Hope:2001p3220,Poulsen:2001p023604,Savage:2006p033620}

The Wigner method, on the other hand, is an approximate method that is valid for large photon number $\overline{n}$ and short fibre length $L$. Here it is  symmetrically ordered correlations that correspond to stochastic averages.  Because of this symmetric ordering, the quantum effects enter via vacuum noise in the initial conditions\cite{Carter:1995p3274}, making it a simple and efficient method for squeezing calculations\cite{Drummond:1993p279}. It is also enjoying increasing utility in BEC simulations\cite{Steel:1998p4824}.

\subsection{Wigner equations}
The Wigner representation maps the operator equations of motion onto (almost) equivalent
stochastic phase-space equations.
The mapping is not exact because the `nonlinear' term leads to higher-order
(higher than second) derivatives in the equation for the Wigner function,
which must be neglected in order that the mapping to stochastic equations
can be completed. These neglected terms are the ones, for instance, which would allow the Wigner function to become negative.  

The resultant equations are, up to a constant phase rotation,
 \begin{eqnarray}
\frac{\partial}{\partial t}\Psi(z,t) & = & \lc-i\sum_{k}R_{k}\left\{ b_{k}+b_{k}^{*}\right\} - v\frac{\partial}{\partial z}\right.\nonumber\\
&&\left.+\frac{i\omega''}{2}\frac{\partial^{2}}{\partial z^{2}}+i\chi^{E}|\Psi|^{2}\rc\Psi\nonumber\\
\frac{\partial}{\partial t}b_{k}(z,t) & = & -i\omega_k b_{k}-iR_{k}\left(|\Psi|^{2}-\frac{1}{2\Delta z}\right),\nonumber\\\end{eqnarray}
where we have assumed that the fields will be discretized over a lattice with segment size $\Delta z$.  
The initial conditions are \begin{eqnarray}
b_{k}(z,0) & = & \Gamma_{k}^{b}(z)\nonumber\\
\Psi(z,0) & = & \ex{\hat{\Psi}(z,0)}+\Gamma_{\Psi}(z),\end{eqnarray}
where the stochastic terms have correlations \begin{eqnarray}
\ex{{\Gamma_{k'}^{b}}^{*}(z')\Gamma_{k}^{b}(z)} & = & \left\{ n_{\textrm{{th}}}(\omega_k)+\frac{1}{2}\right\} \delta_{k,k'}\delta(z-z')\nonumber\\
\ex{\Gamma_{\Psi}^{*}(z')\Gamma_{\Psi}(z)} & = & \frac{\delta(z-z')}{2}.\end{eqnarray}

\subsection{$+P$ equations}
Phase-space equations that correspond exactly to the operator equations can be defined over a doubled phase space using the $+P$ representation.  Quantum effects enter here through multiplicative noise terms in the equations, which generally lead to a larger sampling error than the Wigner method for squeezing calculations.  While the Wigner method was used for nearly all of the simulations presented here, the $+P$ method provides important benchmark results, and was used to check the validity of the Wigner calculations in key cases.

The resultant $+P$ equations are 
\begin{eqnarray}
\frac{\partial}{\partial t}\Psi(z,t) & = & \lc-i\sum_{k}R_{k}\left\{ b_{k}+b_{k}^{+}\right\} \Delta\omega - v\frac{\partial}{\partial z} \right.\nonumber\\
&&\left.+\frac{i\omega''}{2}\frac{\partial^{2}}{\partial z^{2}}+i\chi^{E}\Psi^+\Psi +\sqrt{i}\Gamma^{E} +i\Gamma^{R} \rc\Psi \nonumber\\
\frac{\partial}{\partial t}b_{k}(z,t) & = & -i\omega_k b_{k}-iR_{k}\Psi^+\Psi +\Gamma^{R}_k,
\label{pp1}
\end{eqnarray}
with equations for $\Psi^+$ and $b_k^+$ that have a conjugate form but with some independent noise terms.  The initial conditions are
 \begin{eqnarray}
b_{k}(z,0) & = & \Gamma_{k}^{b}(z)\nonumber\\
\Psi(z,0) & = & \ex{\hat{\Psi}(z,0)},
\label{pp2}
\end{eqnarray}
 where the stochastic terms have correlations \begin{eqnarray}
\ex{{\Gamma_{k'}^{b}}^{+}(z')\Gamma_{k}^{b}(z)} & = & n_{\textrm{{th}}}(\omega_k) \delta_{k,k'} \delta(z-z')\nonumber\\
\ex{\Gamma^{E}(z',t')\Gamma^{E}(z,t)} & = & \chi^{E}\delta(z-z')\delta(t-t')\nonumber\\
&=&\ex{\Gamma^{E+}(z',t')\Gamma^{E+}(z,t)},\nonumber\\
\ex{\Gamma^{R}(z',t')\Gamma^{R}_k(z,t)} & = & R_k\delta(z-z')\delta(t-t')\nonumber\\
&=&\ex{\Gamma^{R+}(z',t')\Gamma^{R+}_k(z,t)},
\label{pp3}
 \end{eqnarray}
with all other correlations zero.

In writing down explicit equations for the phonon variables, we have followed the approach of Carter\cite{Carter:1995p3274}.  In this approach there is some freedom in how the Raman noise is distributed between the photon and phonon variables, a fact which could be exploited to optimise the performance of the simulations.  The alternative approach, as in \cite{Drummond:2001p139}, analytically integrates the phonon variables out, to give nonlocal equations for the photon fields.

\subsection{Scaling}

To simplify the numerical calculation, we transform to a propagating
frame of reference with dimensionless variables: $\tau=(t-z/v)/t_{0}$,
$\zeta=z/z_{0}$ and $\Omega=\omega t_{0}$, where $z_{0}=t_{0}^{2}/k''$.
The fields are also rescaled: $\phi=\Psi\sqrt{vt_{0}/\overline{n}}$
and $\beta_{k}=r_{k}b_{k}\exp(i\Omega\tau)\sqrt{z_{0}/t_{0}\overline{n}}$,
where $r_{k}=R_{k}\sqrt{\overline{n}z_{0}/t_{0}v^{2}}$ is the rescaled
Raman coupling, which is related to the Raman gain $\alpha^{R}(\Omega)$
via $r_{k}=\sqrt{\alpha^{R}(k\Delta\Omega)/2\pi}$.
The quantity $\bar{n}=v^{2}t_{0}/\chi z_{0}$ gives the typical number
of photons in a soliton of width $t_{0}$. The effective nonlinearity
that gives rise to solitons has both electronic and Raman contributions:
$\chi=\chi_{E}+\chi_{R}$, where the Raman contribution is estimated
to be a fraction $f=\chi/\chi_{R}\simeq0.15$ of the total.

For $v^{2}t_{0}^{2}\ll z_{0}^{2}$, the rescaled Wigner equations are:
\begin{eqnarray}
\frac{\partial}{\partial\zeta}\phi(\zeta,\tau) & = & \lc-i\sum_{k}\left\{ \beta_{k}\exp(-i\Omega\tau)+\beta_{k}^{*}\exp(i\Omega\tau)\right\} \Delta\Omega \right.\nonumber\\
&&\left.+\frac{i}{2}\frac{\partial^{2}}{\partial\tau^{2}}+i(1-f)|\phi|^{2}\rc\phi\nonumber\\
\frac{\partial}{\partial\tau}\beta_{k}(\zeta,\tau) & = & -ir_{k}^{2}\left(|\phi|^{2}-\frac{vt_{0}}{2\overline{n}z_{0}\Delta\zeta}\right)\exp(i\Omega\tau),\end{eqnarray}
 with initial conditions \begin{eqnarray}
\beta_{k}(\zeta,\tau & = & -\infty)=\Gamma_{k}^{\beta}(\zeta)\nonumber\\
\phi(\zeta=0,\tau) & = & \sqrt{\frac{vt_{0}}{\overline{n}}}\ex{\hat{\Psi}(0,t_{0}\tau)}+\Gamma_{\phi}(\tau),\end{eqnarray}
 where the stochastic terms have correlations \begin{eqnarray}
\ex{{\Gamma_{k'}^{\beta}}^{*}(\zeta')\Gamma_{k}^{\beta}(\zeta)} & = & \frac{r_{k}^{2}}{\overline{n}}\left\{n_{k}+\frac{1}{2}\right\} \frac{\delta_{k,k'}}{\Delta\Omega}\delta(\zeta-\zeta')\nonumber\\
\ex{\Gamma_{\phi}^{*}(\tau')\Gamma_{\phi}(\tau)} & \simeq & \frac{\delta(\tau-\tau')}{2\overline{n}},\label{eq:scaled_wigner}
\end
{eqnarray}
where $n_{k}=n_{\textrm{{th}}}(k\Delta\Omega/t_{0})$

For numerical convenience, the field is split into two parts - a coherent field that obeys the classical equations of motion, and a difference field that contains the stochastic evolutions.  The equations of motion of each part are given in appendix \ref{split_fields}. 

The rescaled $+P$ equations follow similarly from Eqs (\ref{pp1})-(\ref{pp3}), and are given in appendix \ref{rescaled_pp}.  Because of the much larger sampling error that arises in the $+P$ calculations, we make use of the fact the Wigner method calculates the linearised evolution exactly, and use the $+P$ method only to calculate the difference between the linearised and full evolution.  If $\phi_{WL}$ is a Wigner solution to the linearised equations, and $\phi_{PL}$ and $\phi_{P}$ are $+P$ solutions to the linearised and full equations, respectively, calculated with identical noise sources, then the final solution is $\phi = \phi_{P}-\phi_{PL} + \phi_{WL}$.  Because the difference between the full and linearised solutions is small, $\phi_P$ and $\phi_{PL}$ have very similar fluctuations in a given run; taking the difference removes most of the large $+P$ fluctuations, and adds in only the small Wigner fluctuations.

\section{Outputs and Moments}
\label{Outputs}
We find that good precision (a few percent of the squeezing in decibels) is obtained when averages are calculated using 1000 realisations of the Wigner equations.  For further precision, 10,000 trajectories can be used, in which case we find that the sampling error cannot be distinctly plotted on the graphs.  With the $+P$ method, on the other hand, we find that at least 10,000 trajectories are needed in some cases to produce useful results, even when the differencing method is used.

The observable moments in the polarisation squeezing measurements are integrated intensity measurements and their variances, which are neither simply normally ordered nor symmetrically ordered.  Thus the results of the phase-space simulations must be adjusted for reordering, as we describe below.

In the theoretical description of the system, there are two optical fields, corresponding to the two polarisation modes of the fibre: $\widehat{\Psi}_{x}(t,z)$ and $\widehat{\Psi}_{y}(t,z)$.  To describe the polarisation squeezing, we define integrated Stokes operators, which are a generalisation of Eq.~(\ref{eq_qstokes}):
\begin{eqnarray}
\widehat{S}_{0}\equiv\widehat{N}_{xx}(T)+\widehat{N}_{yy}(T), && \widehat{S}_{1}\equiv\widehat{N}_{xx}(T)-\widehat{N}_{yy}(T),\nonumber \\
\widehat{S}_{2}\equiv\widehat{N}_{xy}(T)+\widehat{N}_{yx}(T), && \widehat{S}_{3}\equiv i\widehat{N}_{yx}(T)-i\widehat{N}_{xy}(T),
\end{eqnarray}
where $T$ is the propagation time down the length of fibre and $\widehat{N}_{\sigma\sigma'}(t)=\int dz\widehat{\Psi}_{\sigma}^{\dagger}(t,z)\widehat{\Psi}_{\sigma'}(t,z)$. After the polarisation rotator, the fields are transformed to 
\begin{eqnarray}
\widehat{\Psi}_{x}'(t,z) &=& \cos(\theta/2)\widehat{\Psi}_{x}(t,z) - i \sin(\theta/2)\widehat{\Psi}_{y}(t,z)\nonumber\\
\widehat{\Psi}_{y}'(t,z) &=& \sin(\theta/2)\widehat{\Psi}_{x}(t,z) + i \cos(\theta/2)\widehat{\Psi}_{y}(t,z),
\end{eqnarray}
which leaves $\widehat{S}_{0}$ unchanged but which transforms $\widehat{S}_{1}$ to
\begin{equation}
\widehat{S}_{\theta}=\cos(\theta)\widehat{S}_{1}+\sin(\theta)\widehat{S}_{2}.
\end{equation}

To calculate that squeezing in $\widehat{S}_{\theta}$, we need to calculate the mean $\ex{\widehat{S}_{\theta}}$ and mean-square $\ex{\widehat{S}_{\theta}^2}$.

\subsection{$+P$ Moments}
For the $+P$ method, stochastic averages of the phase-space variables give normally ordered moments.  Thus the mean $\ex{\widehat{S}_{\theta}}$ can be calculated directly, as it already normally ordered.  The mean square, however, requires a reordering:
\begin{eqnarray}
\ex{\widehat{S}_{\theta}^2}&=&\left< :\left( \cos(\theta)\widehat{S}_{1}+\sin(\theta)\widehat{S}_{2}\right)^2 : \right>+ \ex{\widehat{S}_{0}},
\end{eqnarray}
as shown in appendix \ref{Normal}.

For convenience, we define corresponding stochastic polarisation parameters ${s}_j$,  ${s}_\theta$ in terms of the normalised +P fields: $n_{\sigma\sigma'}(\zeta)\equiv \int d\tau{\phi}_{\sigma}^{+}(\tau,\zeta){\phi}_{\sigma'}(\tau,\zeta)$.  The measured variance can then be written:
\begin{eqnarray}
{\rm var}(\widehat{S}_{\theta})&=& \overline{n}^2 \left( \ex{{s}_{\theta}^2}_{+P}  - \ex{{s}_{\theta}}_{+P}^2+ \frac{1}{\overline n} \ex{{s}_{0}}_{+P}  \right ),
\end{eqnarray}
where $\ex{\dots}_{+P}$ denotes a stochastic average with respect to an ensemble of $+P$ trajectories.  The correction term here corresponds to the shot-noise level of a coherent state (for which 
$\ex{{s}_{\theta}^2}_{+P}  = \ex{{s}_{\theta}}_{+P}^2$): ${\rm var}(\widehat{S}_{\theta})_{\rm coh} = \ex{\widehat S_0} = \overline n  \ex{{s}_{0}}_{+P} $.  Thus the amount of squeezing in decibels is given by:
\begin{equation}
{\rm Squeezing (dB)} = 10 \log \frac{\overline n \ex{{s}_{\theta}^2}_{+P}  - \overline n \ex{{s}_{\theta}}_{+P}^2+ \ex{{s}_{0}}_{+P}}{\ex{{s}_{0}}_{+P}}.
\end{equation}

\subsection{Wigner Moments}
Stochastic averages in the Wigner method correspond to symmetrically ordered products, thus making a reordering necessary for both the mean and variance of the integrated intensity measurements.  First we note the symmetric form of $\widehat{N}_{\sigma\sigma'}$:
\begin{eqnarray}
\left. \widehat N_{\sigma\sigma'} \right|_{\rm sym}  &=&  \frac{1}{2}\int dz\left\{\widehat{\Psi}_{\sigma}^{\dagger}(z)\widehat{\Psi}_{\sigma'}(z) + \widehat{\Psi}_{\sigma'}(z)\widehat{\Psi}_{\sigma}^{\dagger}(z) \right\} \nonumber \\
&=&  \widehat N_{\sigma\sigma'} + \frac{1}{2}\delta_{\sigma\sigma'}M,
\end{eqnarray}
where $M$ is the number of Fourier modes used to decompose the pulse shape.  Because $\widehat S_2$ and $\widehat S_3$ contain only cross-polarisation coherences, there is no correction from reordering.   In $\widehat S_1$, the corrections from horizontal and vertically polarised terms cancel out.  Thus it is only the total intensity that requires a correction, and this corresponds to the vacuum-energy contribution: 
\begin{eqnarray}
\left. \widehat S_0 \right|_{\rm sym}  &=&   \widehat S_0 + M
\end{eqnarray}
 
The variance of the Stokes operators contain terms with products of four operators, each of correspond to 24 possible orderings.  As appendix \ref{Symmetric} shows, most of the corrections cancel out, leaving: 
\begin{eqnarray}
\left .\widehat{S}_{\theta}^2\right|_{\rm sym} &=& \widehat{S}_{\theta}^2 + \frac{1}{2}M.
\end{eqnarray}

Similarly to above, we can define an analogous stochastic polarisation parameter  ${s}_\theta$ in terms of the normalised Wigner fields: $n_{\sigma\sigma'}(\zeta)\equiv \int d\tau{\phi}_{\sigma}^{*}(\tau,\zeta){\phi}_{\sigma'}(\tau,\zeta)$.  The measured variance can then be written:
\begin{eqnarray}
\rm{var}(\widehat{S}_{\theta})&=& \overline{n}^2 \left( \ex{{s}_{\theta}^2}_{W}  - \ex{{s}_{\theta}}_{W}^2- \frac{1}{2\overline n^2} M  \right ),
\end{eqnarray}
where $\ex{\dots}_{W}$ denotes an stochastic average with respect to an ensemble of Wigner trajectories.  The shot-noise reference level is given by ${\rm var}(\widehat{S}_{\theta})_{\rm coh} = \ex{\widehat S_0} = \overline n  \ex{{s}_{0}}_{W} - M$.  Thus the amount of squeezing in decibels is 
\begin{equation}
{\rm Squeezing (dB)} = 10 \log \frac{\overline n \ex{{s}_{\theta}^2}_{W}  - \overline n \ex{{s}_{\theta}}_{W}^2 - \frac{1}{2\overline n} M }{\ex{{s}_{0}}_{W} - \frac{1}{\overline n} M}.
\end{equation}


\section{Experiment}
\label{Experiment}

The laser system used in these experiments is a home-made solid state laser where Cr$^{4+}$:YAG is the active medium~\cite{spaelter97.apb}. This system emits pulses with temporal widths of $\tau_0=$130-150~fs at a central wavelength $\lambda_0$ between 1495-1500~nm. These ultrashort pulses exhibit a bandwidth limited secant-hyperbolic spatial amplitude envelope and are thus assumed to be solitons. The laser repetition rate is 163~MHz and the average output power lies between 60 and 90~mW corresponding to pulse energies of 370-550~pJ. 

\begin{figure}[h]
\begin{center}
\includegraphics[scale=0.5]{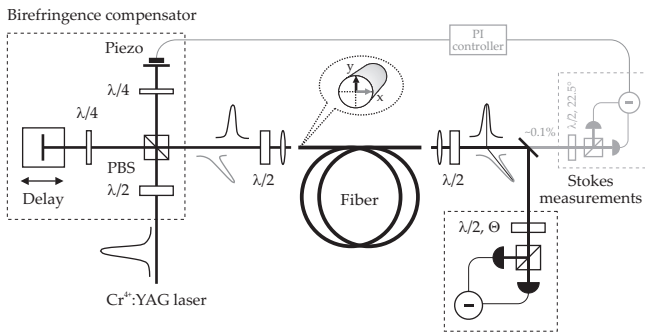}
\caption{\it Schematic of the single-pass method for the efficient production of polarization squeezed states. The Stokes measurement after the fiber scans the dark $\hat{S}_1$-$\hat{S}_2$ plane of the circularly $\langle\hat{S}_3\rangle\neq 0$ polarized output.} \label{fig_polsqsetup}
\end{center}
\end{figure}

In the present configuration, pictured in Fig.~\ref{fig_polsqsetup}, laser pulses are coupled into only one end of the glass fiber. This produces quadrature squeezing rather than amplitude squeezing which is not directly detectable (see Fig.~\ref{fig_kerrsq}(a)). However, overlapping two such independently and simultaneously squeezed pulses after the fiber allows access to this quadrature squeezing by measurement of the Stokes parameters (Fig.~\ref{fig_kerrsq}(b)). This requires the compensation of the fiber birefringence, which we choose to carry out before the fiber to avoid unnecessary losses to the squeezed beams. The optical fiber used was the FS-PM-7811 fiber from 3M, chosen for its high birefringence, i.e. good polarization maintenance, as well as its relatively small mode field diameter, i.e. high effective nonlinearity and thus low soliton energy. The most relevant fiber parameters are listed in Table~\ref{tab_fiber}. 

\begin{table}[h]
	\centering
	\begin{tabular}{|l|c|c|c|c|}
		\hline
		\quad Parameter & Symbol & Fibre I & Fibre II & Units  \\
		\hline
		\hline
		Mode field diameter & {\it d} & \( 5.42\) & \(  5.69 \) & \( \mu \)m  \\
		\hline
		Nonlinear refractive & n\( _{2} \) & 2.9 &2.9 & m \(^{2}\)/W  \\
		index (\( \times 10^{-20}\) ) & & & \\
		\hline
		Effective nonlinearity & \( \gamma \) & 5.3 & 4.8 & 1/(m\(\cdot\)W)  \\
		(\( \times 10^{-3}\) ) & & & \\
		\hline
		Soliton energy & \( E_{Sol} \) & 56 & 60 & pJ  \\
		\hline
		Dispersion & \( k'' (=\beta_{2}) \) & -10.5 & -11.1  & fs\( ^{2} \)/mm  \\
		\hline
		Attenuation @ 1550~nm & \(  \alpha \) &  1.82 & 2.03 & dB/km  \\
		\hline
		Beat length  & \(  L_b \) & 1.67& 1.67 & mm  \\
		\hline
		Polarization crosstalk & \(\Delta P\) & \( < -23\) & \( < -23 \) & dB \\
		per 100m  & & & \\
		\hline
	\end{tabular}\\
\caption{\it Values for the material parameters for the 3M FS-PM-7811 fiber.  Fibres I and II refer to two different production runs. All values (when not otherwise stated) are for $\lambda_0= 1500$~nm and $\tau_0=130$~fs.}
\label{tab_fiber}
\end{table}

For experimental ease, the polarization of the beam after the fiber was set to be circular, e.g. $\sigma_+$. The orthogonal Stokes parameters in the dark $\hat{S}_1$-$\hat{S}_2$ plane, given by Eq.~\ref{rotation}, are measured by rotating a half-wave plate before a polarising beam splitter, as in Fig.~\ref{fig_polsqsetup}.  Equations (\ref{eq_polsq_darkmode}-\ref{eq_polsq_brightmode}) provide an interpretation of the classical excitation in $\hat{a}_{\sigma_+}$ as a perfectly matched local oscillator for the orthogonally polarized dark mode $\hat{a}_{\sigma_-}$. The phase between $\hat{a}_{\sigma_+}$ and $\hat{a}_{\sigma_-}$ varies with the rotation of the half-wave plate angle, $\Theta$, to give the phase-space angle $\theta=4\Theta$. This noise level was compared with the respective Heisenberg limit. The sum photon current, $\hat{S}_0$, gives the amplitude noise of the input beam, for a Kerr-squeezed state this equals the shot noise. This reference level was verified by observation of the balanced homodyne detection of a coherent state as well the sum of the balanced homodyne detection of the $x$- and $y$-polarized modes.

The polarizing beam splitter outputs were detected by two balanced photodetectors based on pin photodiodes. The detectors had a DC output ($<$1~kHz) to monitor the optical power as well as an AC output (5-40~MHz). This frequency window was chosen to avoid low frequency technical noise and the high frequency laser repetition rate. The sum and difference of the detectors' AC photocurrents, representing the noise of different Stokes variables, were fed into a spectrum analyzer (Hewlett-Packard 8595E) to measure the spectral power density at 17.5~MHz with a resolution bandwidth of 300~kHz and a video bandwidth of 30~Hz. 

\section{Results -  experiment and simulation}
\label{Results}

\subsection{Characterising the single-pass method}

\begin{figure}[h]
\centering
\includegraphics[width=8.5cm]{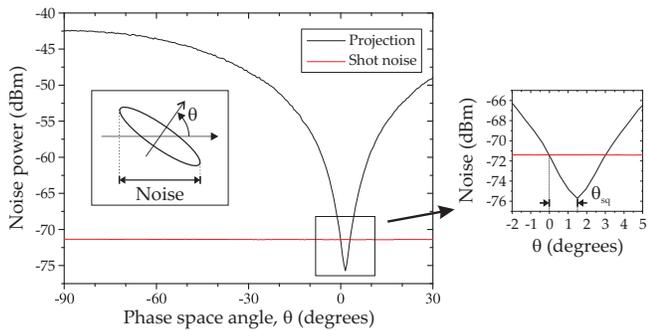}
\caption{\it Noise power against phase-space rotation angle for the rotation of the measurement half-wave plate for a pulse energy of 83.7~pJ using 13.3~m of fiber I. Inset: Schematic of the projection principle for angle $\theta$.} \label{fig_polsq_rotation}
\end{figure}

The single-pass squeezing method allows the measurement of greater squeezing as well as the direct and full characterization of the bright Kerr-squeezed beams~\cite{heersink05.ol, Marquardt:2007p220401}. Both of these traits are visible in Fig.~\ref{fig_polsq_rotation}. Here the measured AC noise as a function of the rotation of a half-wave plate (by the angle $\Theta$) in the dark Stokes plane is seen. A progression between very large noise and squeezing is observed, as expected from the rotation of a fiber squeezed state. Plotted on the x-axis is the projection angle $\theta$, i.e. the angle by which the state has been rotated in phase space, inferred from the wave plate angle ($\theta=4\Theta$). Here pulses of 83.7~pJ were transmitted through 13.3~m of optical fiber and the electronic signals were corrected for the $-86.1\pm0.1$~dBm dark noise.

For $\theta=0$, an $\hat{S}_1$ measurement gives a noise value equal to the shot noise. This corresponds to the amplitude quadrature of the Kerr-squeezed states emerging from the fiber. Rotation of the state by $\theta_{sq}$ makes the state's squeezing observable by projection onto the minor axis of the noise ellipse. Further rotation brings a rapid increase in noise as the excess phase noise, composed of the antisqueezing and the classical thermal noise arising from GAWBS, becomes visible. The maximum noise is observed for $\theta=\theta_{sq}+\frac{\pi}{2}$. Under the assumption of statistically identical but uncorrelated Kerr-squeezed states, this measurement is equivalent to the characterization of the individual squeezed states using standard local oscillator and homodyne detection methods. However here no stabilization is needed after production of the polarization squeezed state. This is advantageous for experiments with long acquisition times, i.e. state tomography, and has indeed allowed the reconstruction of the Wigner function of the dark Stokes plane or Kerr-squeezed states~\cite{Marquardt:2007p220401}. 

\begin{figure}[h]
\centering
\includegraphics[width=7cm]{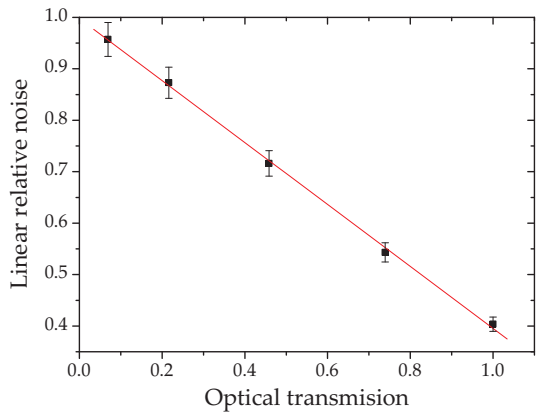}
\caption{\it Linear noise reduction against optical transmission for the polarization squeezing generated by pulses of an energy of 81~pJ in 3.9~m of fiber I.} \label{fig_polsq_attntest}
\end{figure}

It is crucial to ensure that the measured squeezing did not arise from detector saturation or any other spurious effect. This was accomplished observing the noise of a variably attenuated squeezed beam, where a plot of the linear relative noise against the transmission should be linear for true squeezing. A representative plot for a 81~pJ pulse in a 3.9~m fiber exhibiting $-3.9\pm0.3$~dB of squeezing is shown in Fig.~\ref{fig_polsq_attntest}; the linear result is indicative of genuine squeezing.

\begin{figure}[h!]
\centering
\includegraphics[width=7cm]{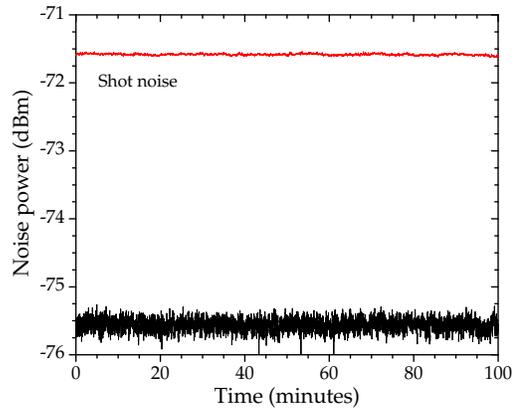}
\caption{\it Plot showing a stable squeezing of $-4.0\pm0.3$~dB over 100 minutes. A 30~m optical fiber with a pulse energy of 80~pJ was used.}
\label{fig_polsq_longsq}
\end{figure}

The single-pass polarization squeezer exhibits a good temporal stability, highlighted by the results in Fig.~\ref{fig_polsq_longsq}. Here the sum (shot noise) and difference (polarization squeezing) channels have been plotted. An average of {-4.0~dB} of squeezing corrected for $-85.8\pm0.1$~dBm of dark noise was measured over 100~minutes. The squeezer used 30~m of optical fiber into which two orthogonally polarized pulses of 40~pJ each were coupled. The most sensitive factor in this setup is the locking of the birefringence compensator. Further important parameters are the coupling of light into the fiber and the laser power stability. All of these parameters are easily held stable by exploiting commercially available components.

\subsection{Squeezing Results}
\begin{figure}[h!]
  \centering
    \includegraphics[width = 7cm]{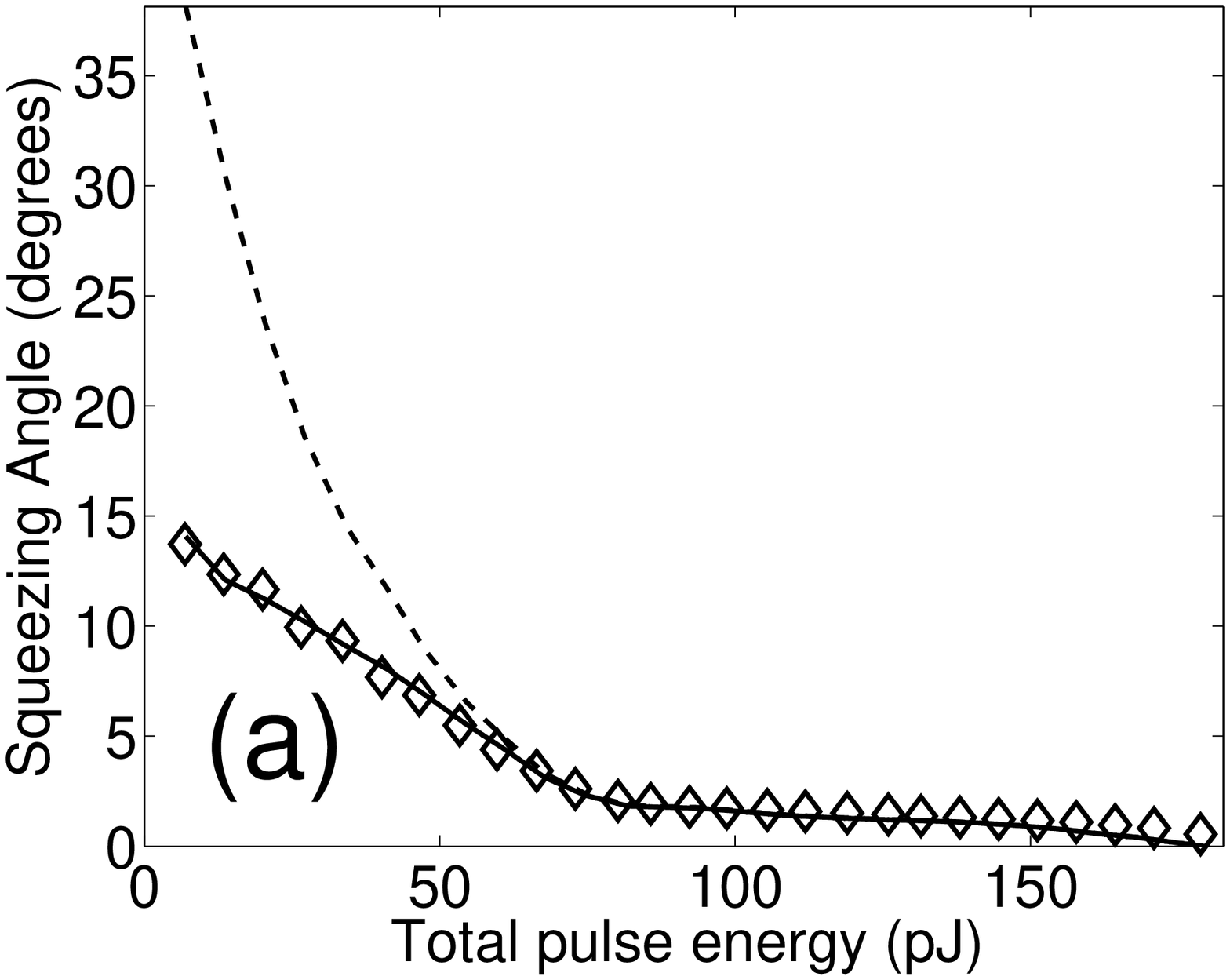}
    \includegraphics[width = 7cm]{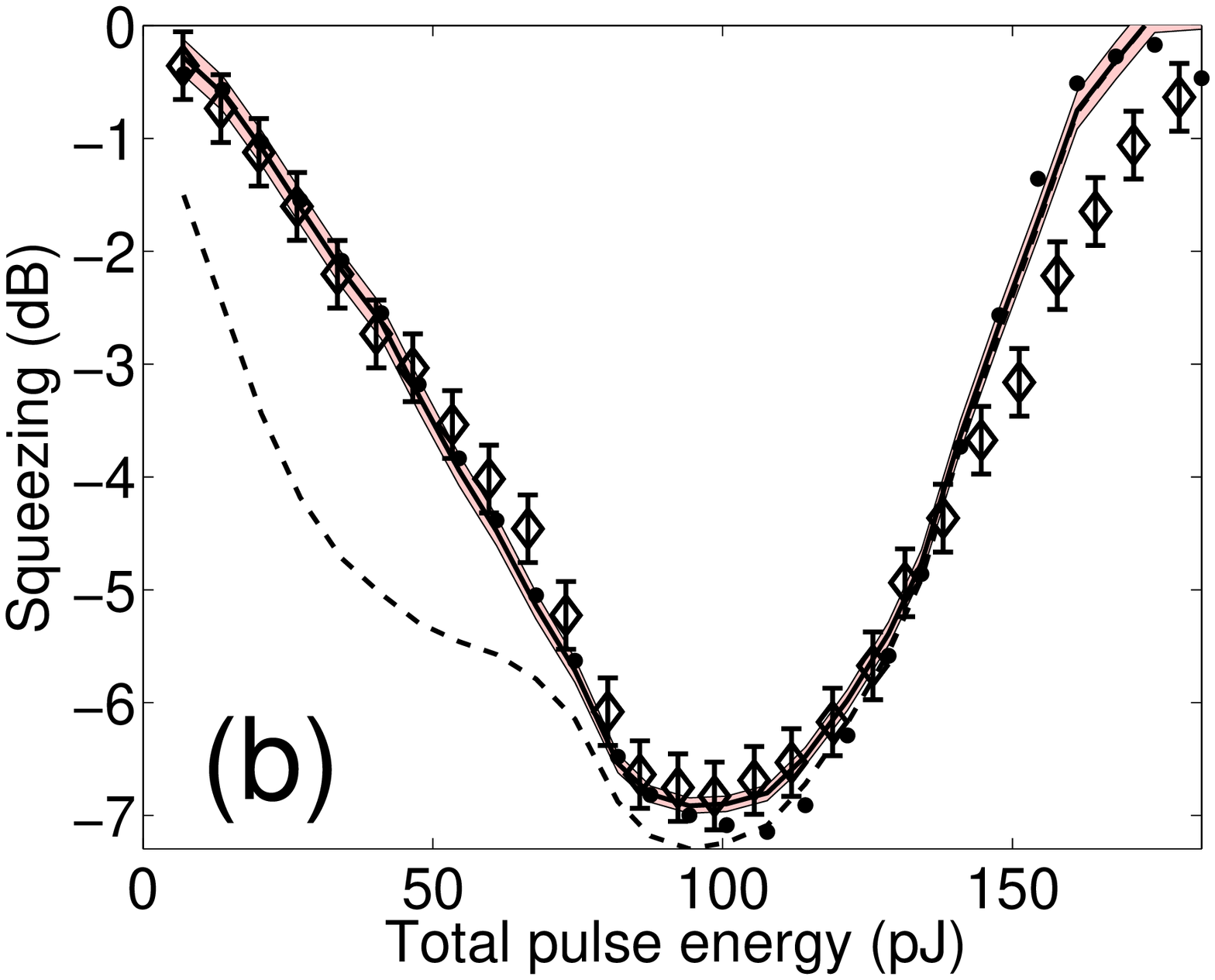}
    \includegraphics[width = 7cm]{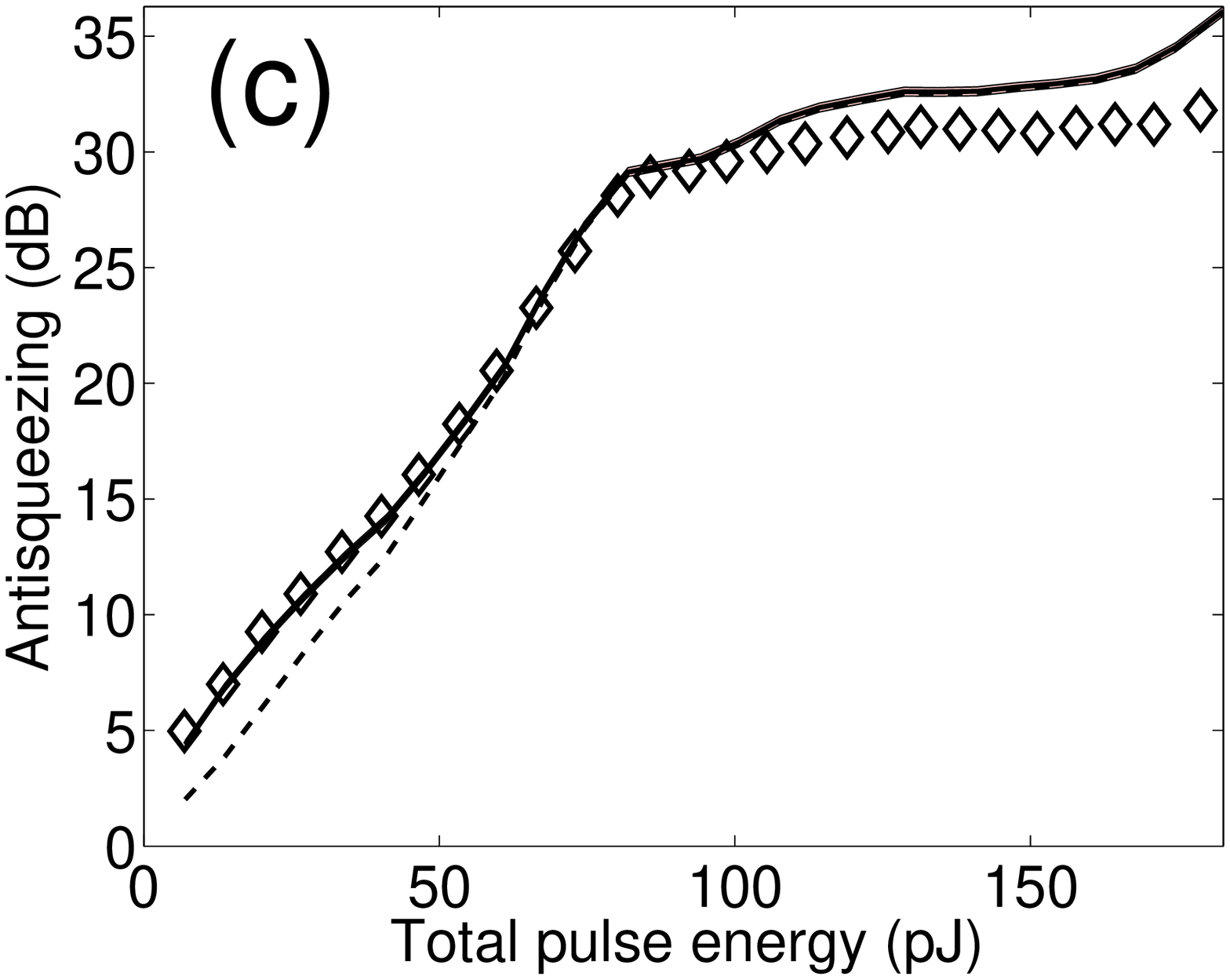}
	\caption{{\it Measurement results and theoretical simulations for 13.2~m 3M FS-PM-7811 fiber (run II) as a function of pulse energy: (a) the squeezing angle, (b) the squeezing and (c) antisqueezing noise. Solid and dashed lines show the simulation results with and without additional phase noise, with linear losses are taken to be 13\%. The shading indicates simulation uncertainly. The simulation result without third-order dispersion is given by the dots in (b).  Diamonds represent the experimental results, with experimental uncertainty indicated by the error bars in the squeezing.  Both the simulation and the experimental errors were too small to be plotted distinctly for the squeezing angle and antisqueezing. The measured noises are corrected for $-85.1\pm0.1$~dBm electronic noise.}}
	\label{fig:results_13.2m}
\end{figure}

The squeezing angle, $\theta_{sq}$ and the squeezed and anti-squeezed quadratures were experimentally investigated as a function of pulse energy from 3.5~pJ to 178.8~pJ, as plotted in (a), (b) and (c), respectively, of Fig.~\ref{fig:results_13.2m} (triangles).  The x-axis shows the total pulse energy, i.e. the sum of the two orthogonally polarized pulses comprising the polarization squeezed pulse.  We observe maximum  squeezing $-6.8\pm0.3$~dB at an energy of 98.6~pJ. The corresponding antisqueezing of this state is $29.6\pm0.3$~dB and the squeezing angle is $1.71^{o}$.  As the optical energy goes beyond 98.6~pJ, the squeezing is reduced, eventually reaching the shot noise limit (SNL), and the increment of antisqueezing slows down to a plateau area. 

The loss of the set-up was found to be 13\%: 5\% from the fibre end, 4.6\% from optical elments and from the fibre attenuation (2.03 dB/km at 1550 nm), 2\% from incomplete intereference between the two polarisation modes ($~99\%$ visibility was measured), and 2\% from the photodiodes.  Thus we infer a maximum polarisation squeezing of $-10.4\pm 0.8$dB.  The improvement in squeezing over previous implementations\cite{heersink05.ol, Corney:2006p023606} of the single-pass scheme is largely due to the low loss achieved here. 

The theoretical simulations for the squeezing, antisqueezing and squeezing angle at different input energies are also given in Fig.~\ref{fig:results_13.2m} by solid and dashed lines. As described in further detail below, the effect of excess phase noise, such as GAWBS~\cite{Corney:2006p023606},  is included by a single-parameter fit between the simulation and experimental data for squeezing angle (shown by solid line in Fig.~\ref{fig:results_13.2m}(a)). The theoretical results for squeezing and antisqueezing then show very good agreement with the experimental data, and are consistent with the measured linear losses of 13\%. From the simulations, the effect of the GAWBS is seen to be a reduction in squeezing for lower energy pulses; above about 100pJ, it has virtually no effect on the squeezing.   Although some deviations appear at very high input energy, the simulations also show the same deterioration of squeezing for higher pulse energies as is seen in the experimental results; this effect does not occur in the simulations if Raman terms are neglected, as we discuss below.

\subsection{Phase-noise and GAWBS}
\label{sec:phasenoise}
Excess phase noise, caused for example by depolarising GAWBS in the fibre, is determined for each fibre length by a single-parameter fit of the experimental and simulation squeezing angles.  We model this by independent random fluctuations in the refractive index at each point along the fibre length.  The cumulative effect on each pulse at a given propagation length is a random phase shift whose variance is proportional to the time-width of the pulse: 
\begin{equation}
\phi(\tau,\zeta) = \phi_0(\tau,\zeta) e^{i\eta},
\end{equation}
where $\ex{\eta^2}   \propto t_0$.

Such phase fluctuations do not affect the number difference measurement $\widehat S_1$, but they do lead to fluctuations in $\widehat S_2$ 
\begin{equation}
\widehat S_2 - \left< \widehat S_2 \right> \simeq  2\eta \overline n \int  |\phi_0(\tau,\zeta)|^2d\tau \propto \eta E 
\end{equation}
where $\eta$ is now taken to describe the relative (depolarising) relative phase shifts.
Thus the variance relative to shot noise of $\widehat S_2$ caused by phase fluctuations scales linearly with the energy of the pulse:
\begin{eqnarray}
\sigma \equiv \frac{ {\rm var} (\widehat S_2 )}{\ex{\widehat{S}_0}} &\propto& \frac{\ex{\eta^2} E^2}{E}  \nonumber \\
&=& c E
\end{eqnarray}
 where the constant of proportionality $c$ is to be determined by the fit.  Here we have assumed that the pulse width is a constant, independent of input energy.  This assumption is not entirely accurate, because unless the energy is the soliton energy for that pulse width, the pulse will reshape to form a soliton, thereby altering the pulse width.   However, for short fibre lengths, this effect should be small, and so we neglect it in our calculations.
 

The effect of the phase noise will be to stretch the squeezing ellipse in the $S_2$ direction, according to the formula:
\begin{equation}\frac{\rm var (\widehat S_\theta)}{\ex{\widehat S_0}} =  a\cos^2(\theta-\theta_K) + b \sin^2 (\theta-\theta_K) + cE\sin^2(\theta), \label{ellipse}
\end{equation}
where $\theta_K$ is the predicted angle from the Kerr-only squeezing, $a$ is the relative Kerr squeezing and $b$ is the relative Kerr anti-squeezing.  These parameters are calculated by the simulation, and are a function of the input energy $E$.  The value of $c$ is determined by fitting the predicted angle of maximum squeezing as a function of $E$ against the observed values.  The predicted angle is obtained from the extrema of the expression in Eq.~(\ref{ellipse}) and the fit is performed with a nonlinear least squares method.  Once the value of $c$ is determined, new values of squeezing and antisqueezing are calculated from eq. (\ref{ellipse}).

\begin{figure}[h!]
\centering
\includegraphics[width = 7cm]{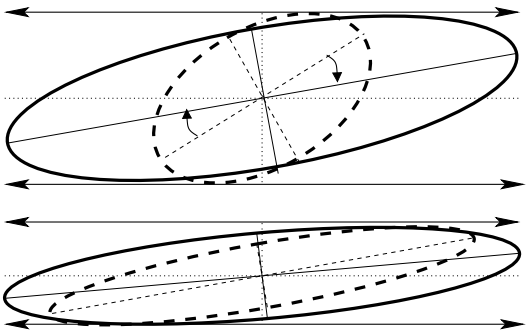}
\caption{\it Illustration of the effect of excess phase noise on different squeezing ellipses. Dashed line gives the Kerr-squeezed ellispse, and solid line gives the ellipse with added phase noise.  The effect on the squeezing and the angle is less for the ellipse with larger Kerr squeezing (lower ellipse).}
 \label{angles}
  \end{figure}

As figure \ref{angles} illustrates, the excess phase noise has a substantial effect, on both the squeezing angle and the amount of squeezing, only at low levels of squeezing.  For highly squeezed light, the Kerr-squeezed ellipse is more closely aligned to the phase quadrature, and thus the phase-noise merely has the effect of increasing the antisqueezing.  This view is confirmed by the results in Fig.~(\ref{fig:results_13.2m}), where the difference between the curves with and without the phase-noise-fitting is discernible only at lower input energies.

\subsection{Third-order Dispersion}

The comparison between theory and experiment confirms the deterioration of squeezing at large pulse energy caused by Raman effects in the fibre.  However there is still some residual discrepancy between theory and experiments, which could be caused by various higher-order effects not included in the theoretical model.  We here explore the effect of third-order dispersion, and find that it accounts for some of the unexplained difference at high energies.

\begin{figure}[h!]
\centering
\includegraphics[width=7cm]{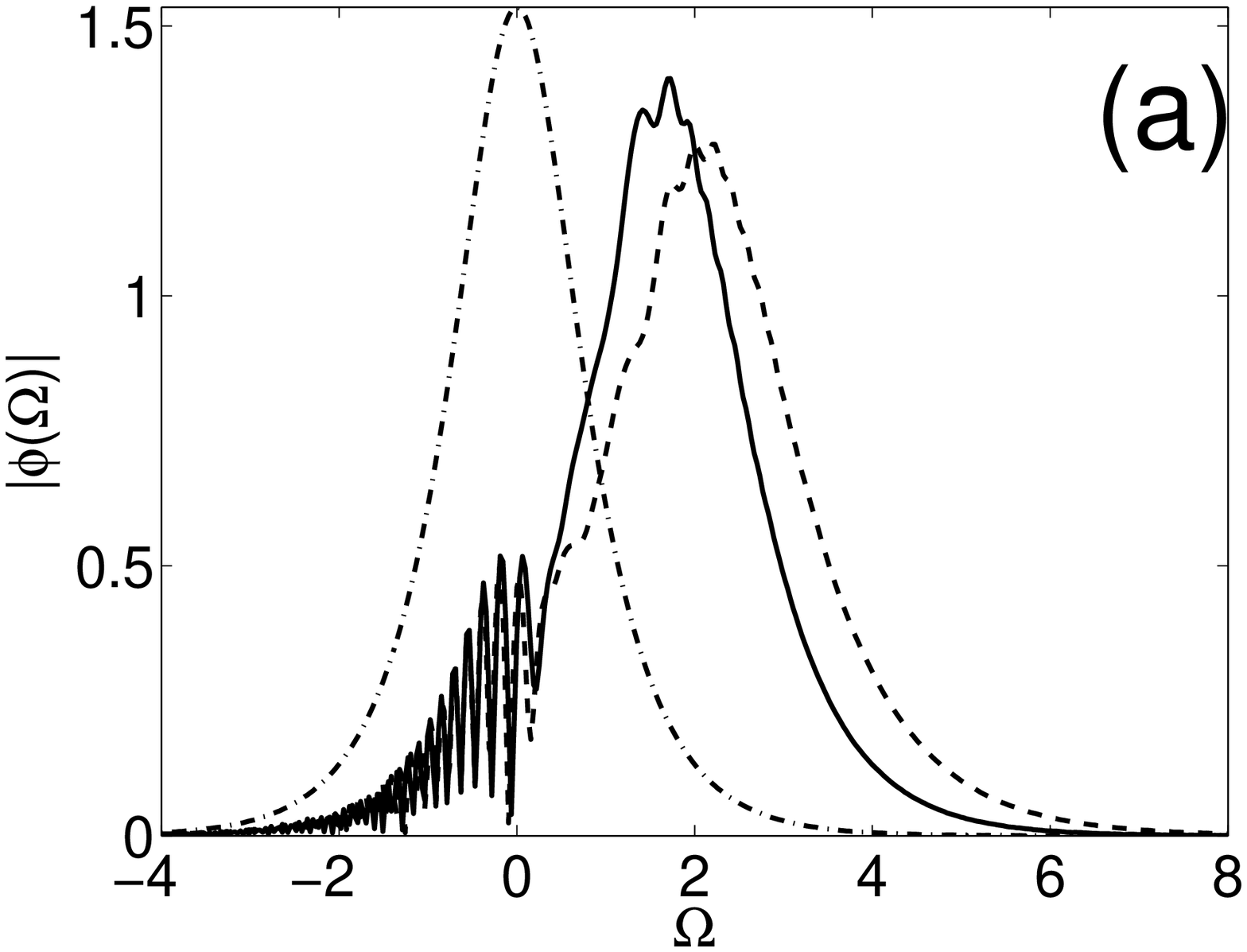}
\includegraphics[width=7cm]{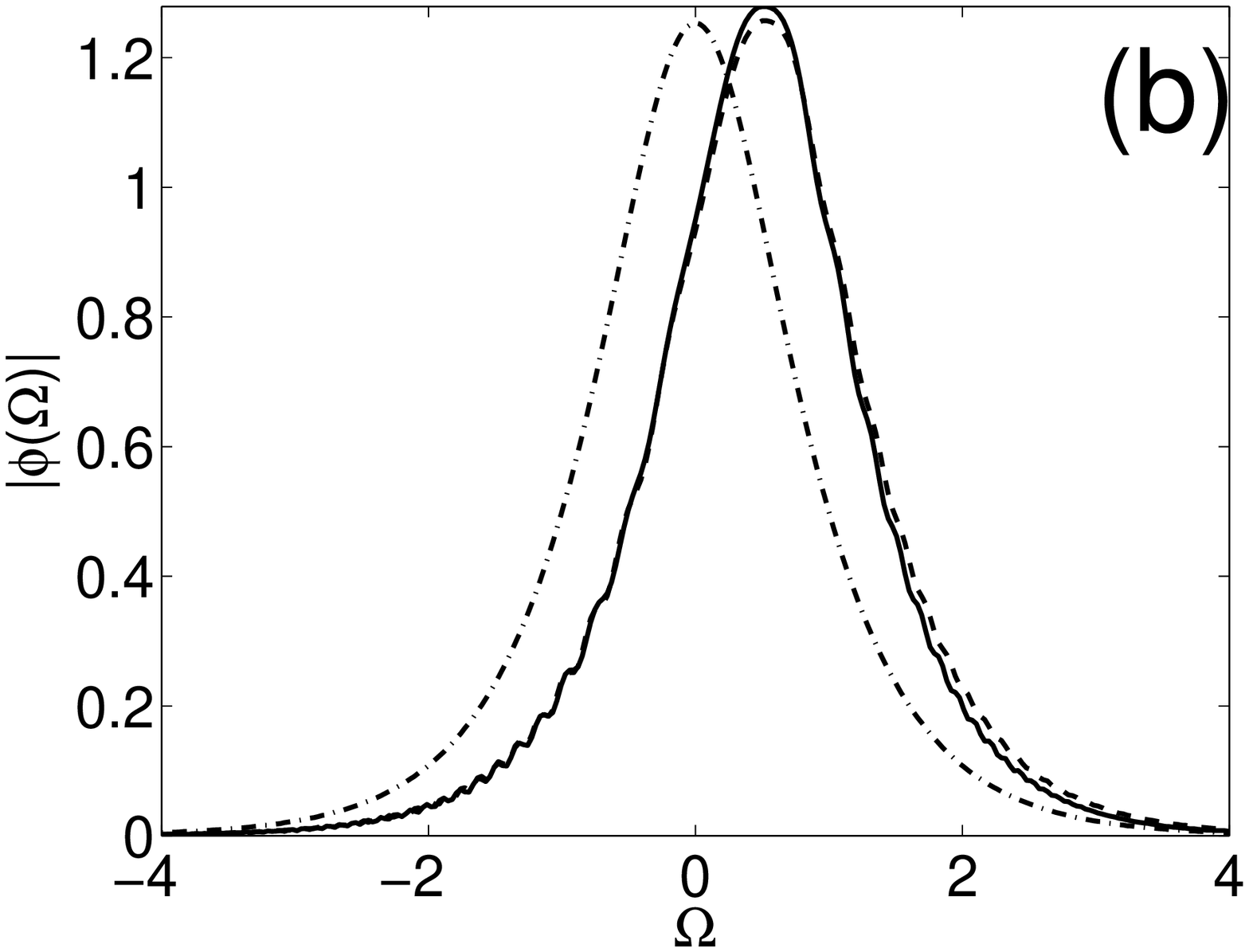}
\includegraphics[width=7cm]{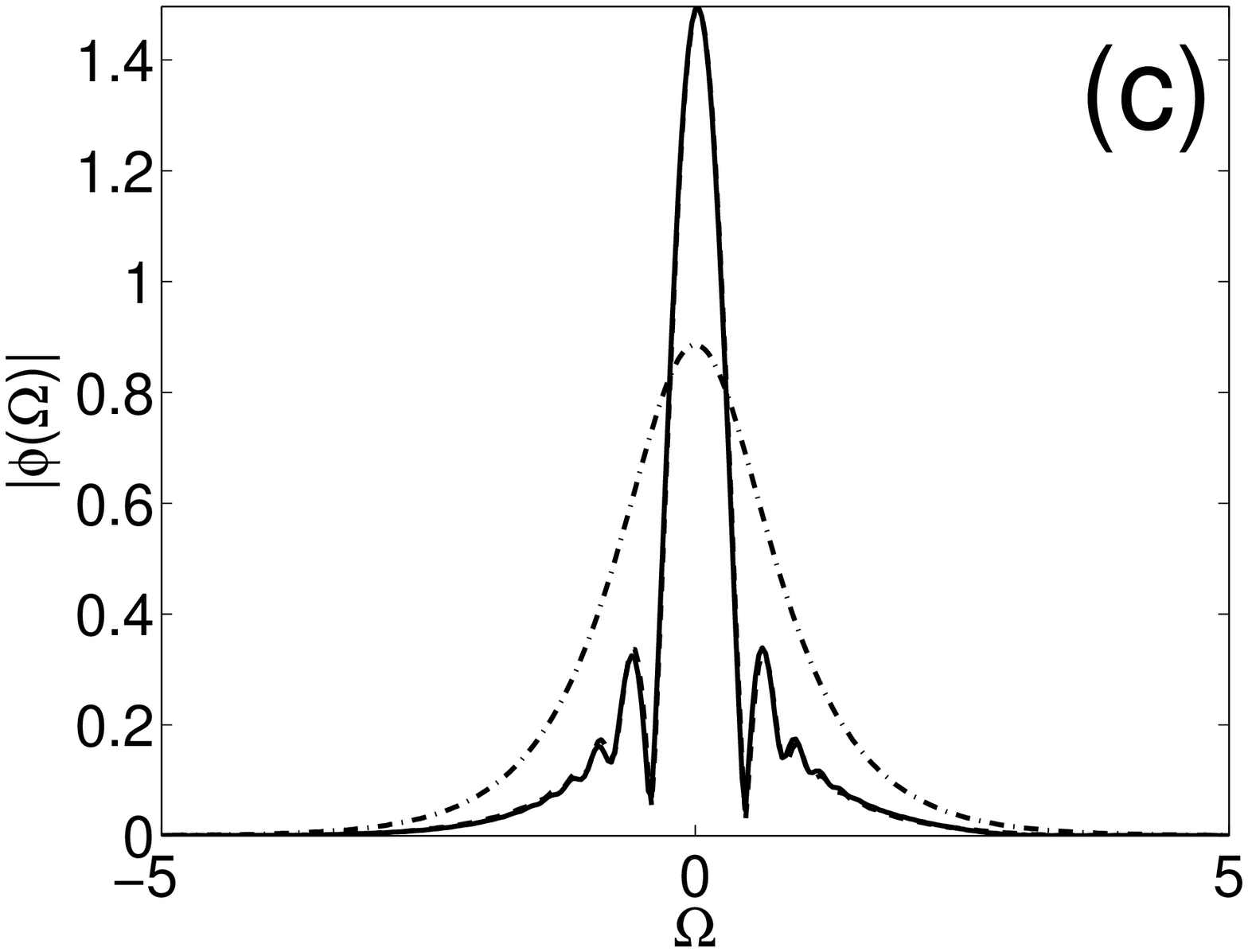}
\caption{\it Simulation pulse-spectrum at pulse energies (a) $1.5E_s$, (b) $E_s$, (c) $0.5E_s$ at rescaled distances of $\zeta = 0$ (dot-dashed), and $\zeta = 25$ with (solid) and without (dashed) third-order dispersion}
\label{TOD}
\end{figure}

Third-order dispersion\cite{Yu:1995p2340} arises from the rate of change of curvature of the dispersion.  It becomes more important for shorter pulses or when operating near the zero-dispersion wavelength\cite{Agrawal:2007}.  In the propagation equation, it appears as an extra term in the scaled equations:
\begin{equation}
\frac{\partial}{\partial \zeta} (\zeta,\tau) = - \frac{B_3}{6} \frac{\partial^3}{\partial\tau^3},
\end{equation}
where $B_3 = k''' z_0/t_0^3$ is a dimensionless third-order dispersion parameter.  For the fibre used in the experiment, the third-order dispersion at $\lambda = 1499$nm is $k''' = 8.38\times 10^{-41} $s$^3$/m, giving $B_3 = 0.097$.  The effect of third-order dispersion on the pulse spectrum for various energies is shown in Fig. \ref{TOD}, where significant differences appear only above the soliton energy.

Third-order dispersion does not have an observable effect on the squeezing angles or the amount of antisqueezing, but its effect can be seen on the squeezing, as shown in Fig.~\ref{fig:results_13.2m}(b) by the difference between the solid and dot-dashed lines.  Below the soliton energy, the third-order dispersion has no observable effect.  It diminishes the amount of squeezing at around the soliton energy, and at higher energies it changes the rate at which squeezing deteriorates as a function of energy.   Far above the soliton energy, there remains some difference between simulation and experiment, which indicates that other higher-order processes may be playing a role at these energies. Because, in any case, the effect of third-order dispersion is rather small,  we do not include it in the other simulation results shown in this paper

\subsection{Raman noise effects}

\begin{figure}[h!]
\centering
\includegraphics[width = 7cm]{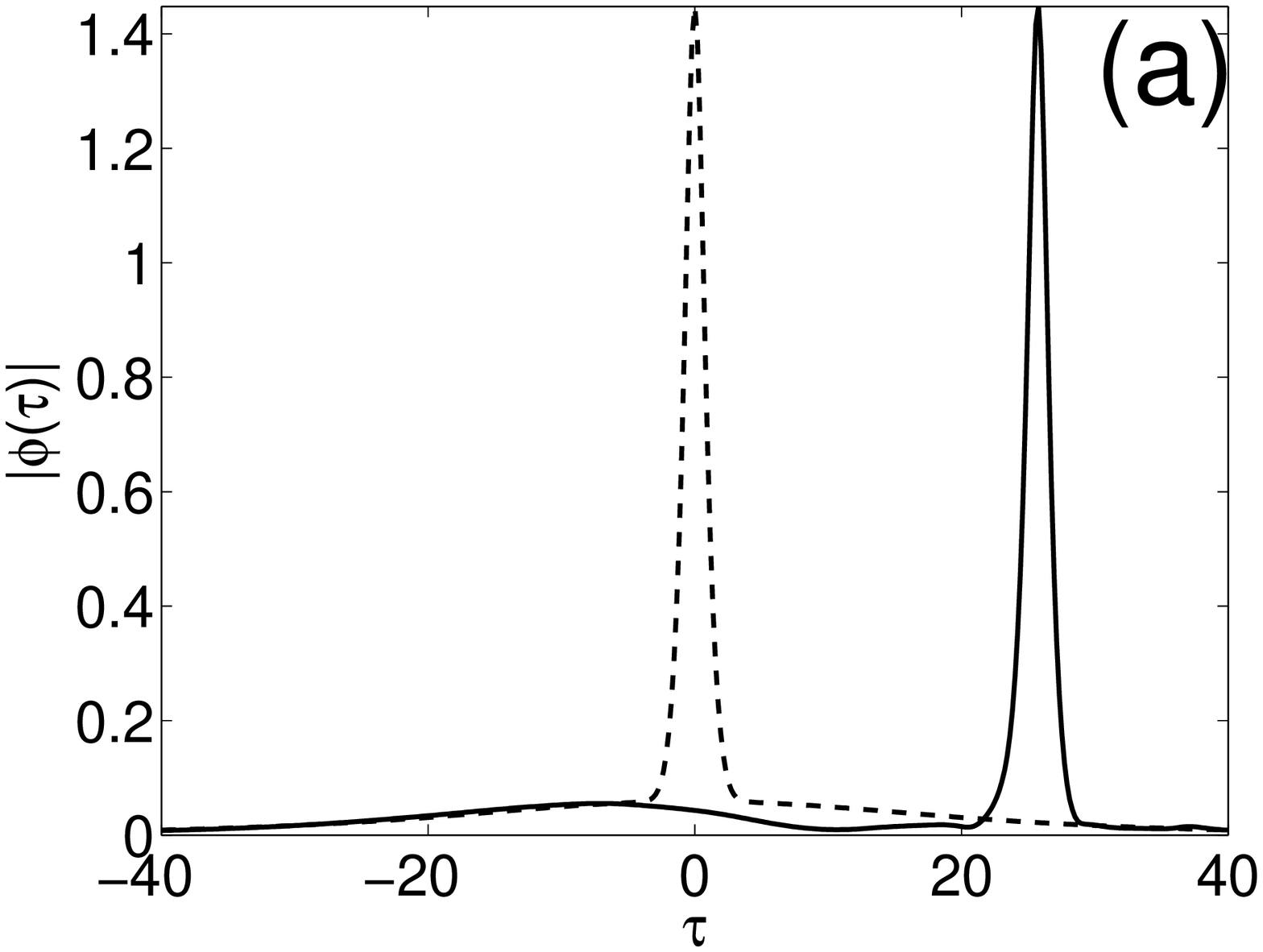}
\includegraphics[width = 7cm]{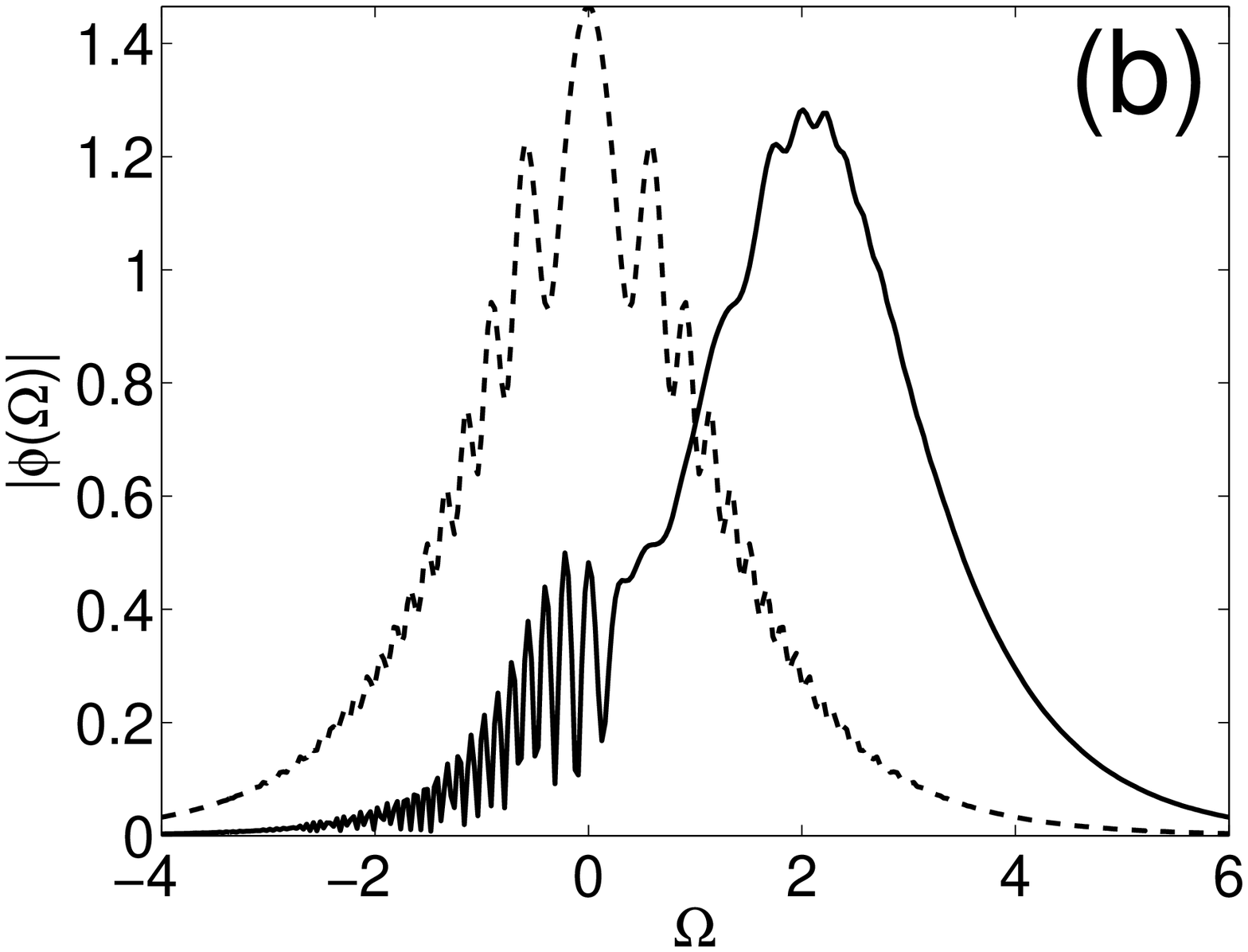}
\caption{ Simulation pulse shape  (a) and spectrum (b) at pulse 
energies $1.5E_s$ and normalised propagation length $\zeta = 25$, with (solid) and without 
(dashed) Raman effects.
}
\label{Ramanpulse}
\end{figure}

The Raman effect has a significant effect on the pulse shape and spectrum for the more intense pulses at these subpicosecond pulse widths.  For a soliton pulse, the effect of the Raman interactions is to induce a frequency shift in the soliton and hence a delay in its arrival time\cite{mitschke86.ol,gordon86.ol}.  For pulses above the soliton energy, the Raman interaction affects the way the pulse reshapes.  With a purely electronic (instantaneous) nonlinearity, the pulse reshapes into a narrower soliton, at the same time shedding radiation that forms a low pedestal underneath the soliton.  In the frequency domain, this results in a modulation of the pulse spectrum.  As Fig.~\ref{Ramanpulse} shows, with the Raman terms included, the reformed soliton separates from the pedestal, which distorts the spectrum into an asymmetric shape. 

\begin{figure}[h!]
\centering
\includegraphics[width = 7cm]{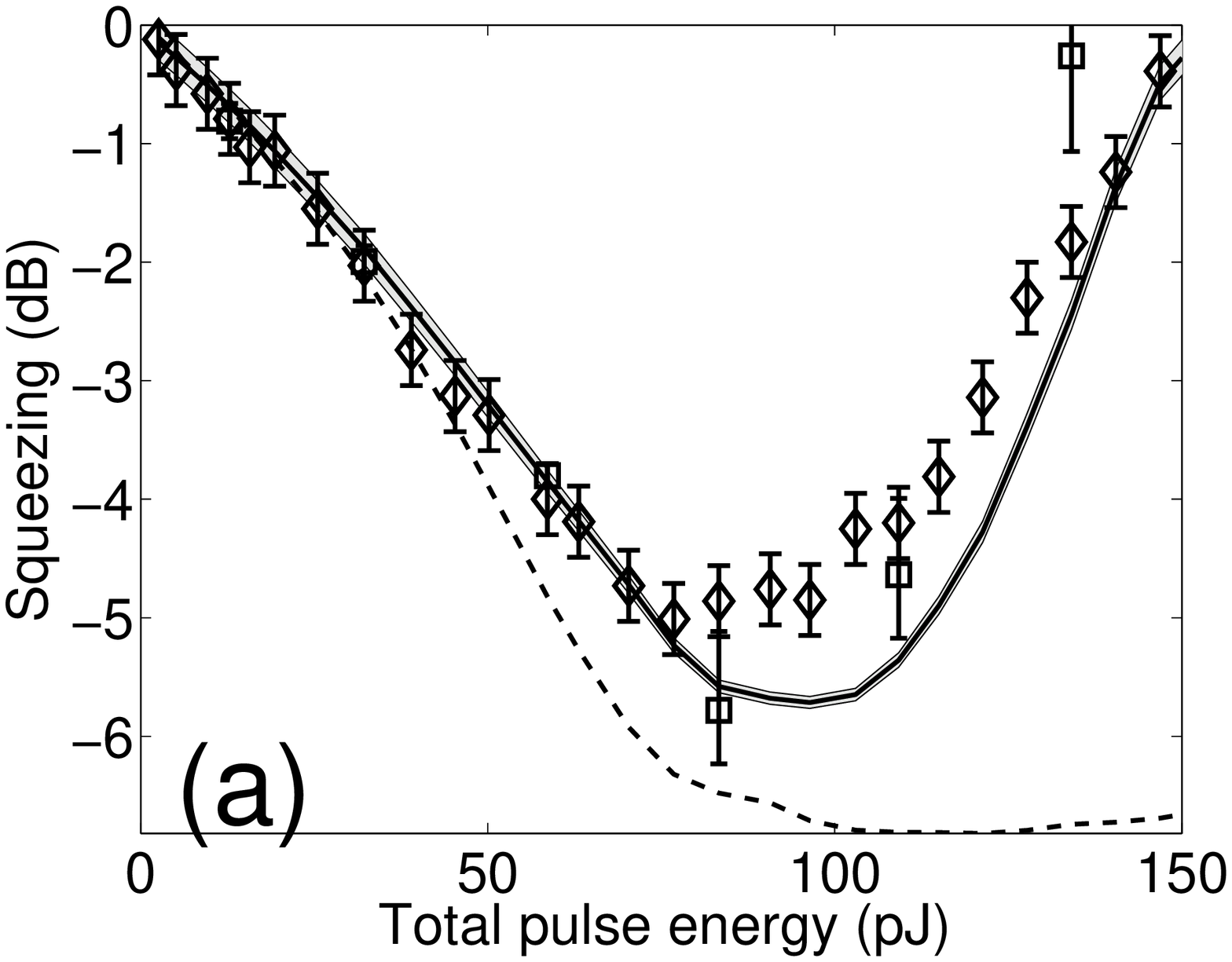}
\includegraphics[width = 7cm]{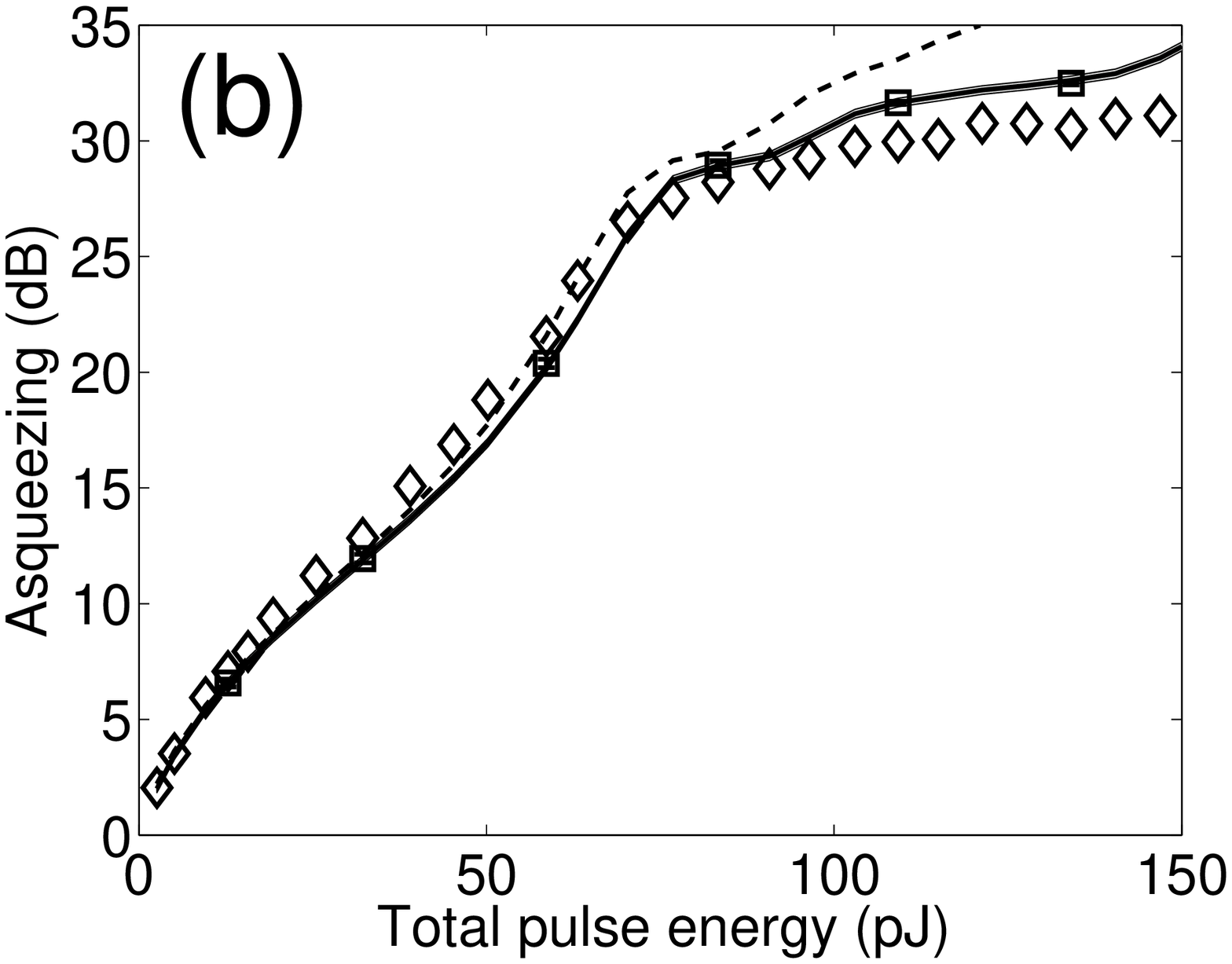}
\caption{Squeezing degradation at high intensity: (a) squeezing and (b) antisqueezing measurements for $L=13.35 \textrm{m}$ of fibre I. Solid and dashed lines shows the simulation results with and without Raman effects (i.e. a purely electronic nonlinearity), respectively.  The data points marked by squares are the results of exact $+P$ calculations with error-bars indicating estimated sampling error. The points at 109~pJ and 134pJ were calculated with 10,000 trajectories; the other 4 $+P$ points were calculated with 1000 trajectories. Note that these results were obtain in an experimental set-up with higher losses than that of Fig.~\ref{fig:results_13.2m}, giving a reduced magnitude of raw squeezing.  The simulations here assume 19.9\% loss} \label{raman_comparison}
\end{figure}

For a pure Kerr effect, the squeezing is proportional to the intensity of the light, which in our case corresponds to the input energy of the pulse.  However the experimental and simulation results clearly show that  while the squeezing increases with input energy over a range of energies, there is a point beyond which the squeezing deteriorates.   This deterioration is largely due to Raman effects, as Fig.~\ref{raman_comparison} reveals, which compares the simulations with and without Raman effects.  In the latter case the nonlinearity is taken to be of the same magnitude as the former but is instantaneous.  Without Raman effects, the squeezing does not suffer the same catastrophic reduction at high energies, but it does appear to saturate at around the soliton energy (2$\times$54pJ), demonstrating that pulse-reshaping effects are also in play.

For  $L=13.35 \textrm{m}$, the optimum energy is around 80\% of the soliton energy.

\subsection{Comparison with exact $+P$ results}
Nearly all of the simulation results presented in this paper were calculated with the truncated Wigner phase-space method, because results can be obtained quickly and with low sampling error.  However, the Wigner technique only provides an approximation to the true quantum dynamics.  While the approximation is usually a good one for intense optical pulses, some deviations from the exact result could in principle occur for long simulation times, or when highly-squeezed states are being produced.  To test the Wigner method, we compared selected points with $+P$ calculations,  and found agreement within the statistical uncertainty.  One example is shown in Fig.~\ref{raman_comparison}, where the $+P$ results are shown as the squares.  As the error-bar indicates, the sampling error for the $+P$ is much larger than that of the Wigner for the more intense pulses, even though 10 times as many trajectories were used for the $+P$ calculation.  Even for the same number of trajectories, $+P$ calculation is more computationally exacting.  This combination of greater computational cost per trajectory and the larger number of trajectories required for a meaningful $+P$ result is why the Wigner technique has been our method of choice for squeezing calculations.  The $+P$ method comes into its own when more exotic quantum states or fewer photons are involved, i.e. when the Wigner technique is not expected to be reliable.  It is also possible that appropriate diffusion\cite{Plimak2001} or drift\cite{Deuar:2006p6847} gauges may improve the performance of the $+P$ calculations.

\subsection{Comparison for different fibre lengths}

\begin{figure}[h!]
\centering
\includegraphics[width = 8cm]{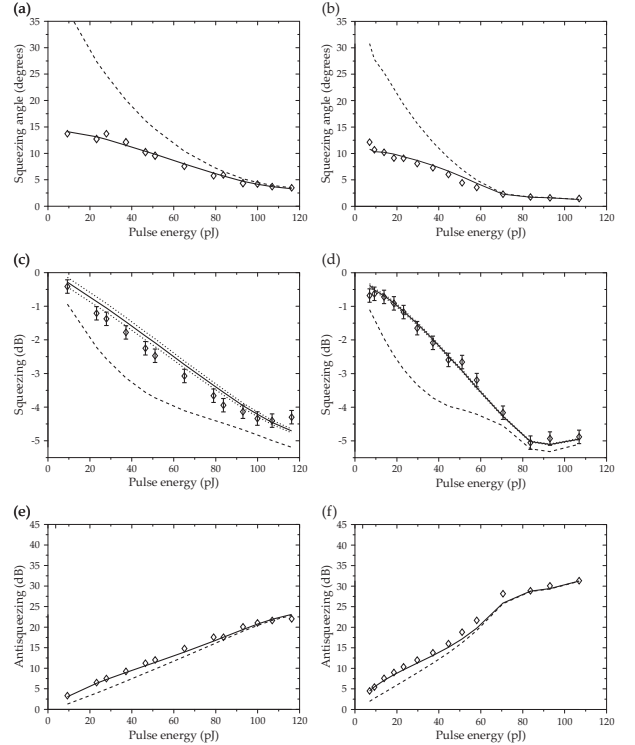}
\caption{\it Experiments (corrected for dark noise) and simulations (with and without fitted phase noise) of the polarization squeezing, antisqueezing and squeezing angle for 3.9 (a, c, e) and 13.3~m (b, d, f) of fiber I} \label{fig_polsq_3.9_13.4m}
\end{figure}

\begin{figure}[h!]
\centering
\includegraphics[width = 8.5cm]{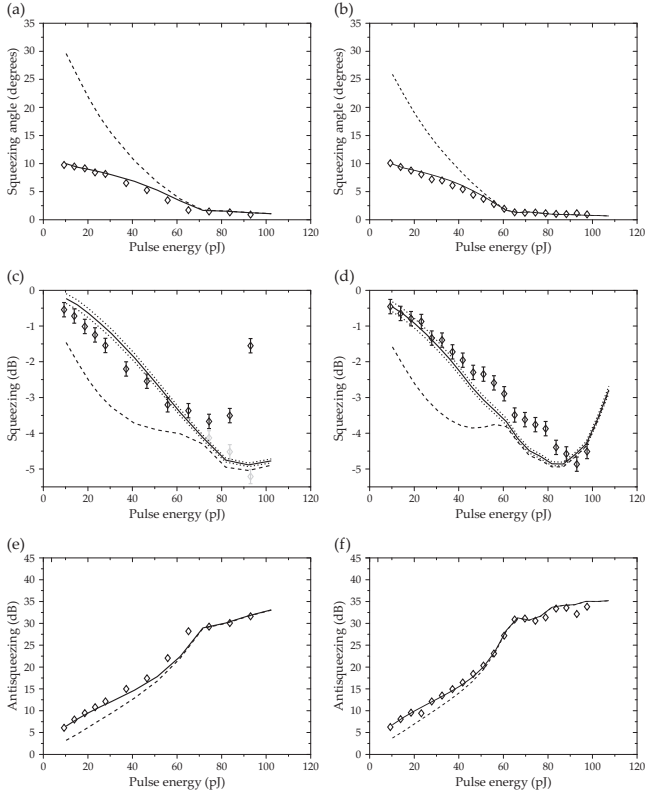}
\caption{\it Experiments (corrected for dark noise) and simulations (with and without fitted phase noise) of the polarization squeezing, antisqueezing and squeezing angle for 20 (a, c, e) and 30~m (b, d, f) of fiber II.  The lighter data points in (c) are from a corrected experimental run.} \label{fig_polsq_20_30m}
\end{figure}

\begin{figure}[h!]
\centering
\includegraphics[width = 8.5cm]{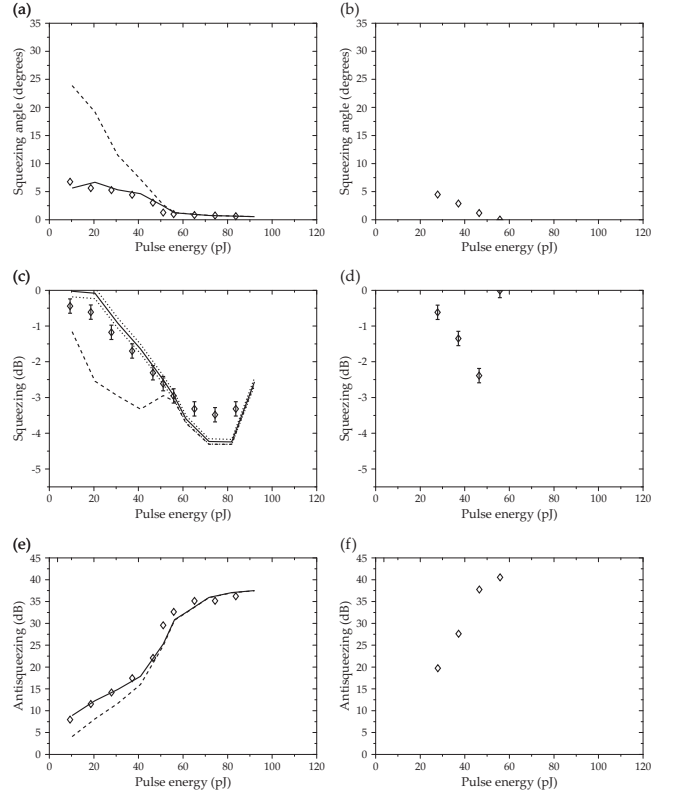}
\caption{\it Experiments (corrected for dark noise) and simulations (with and without fitted phase noise) of the polarization squeezing, antisqueezing and squeezing angle for 50 (a, c, e) and 166~m (b, d, f) of fiber II.  The amount of dispersion over 166~m makes the simulations impractical for this case.} \label{fig_polsq_50_166m}
\end{figure}

The squeezed and antisqueezed quadratures as well as the squeezing angle $\theta_{sq}$ of such polarization states were investigated as a function of pulse energy for different lengths of 3M FS-PM-7811 fiber, as shown in Figs.~\ref{fig_polsq_3.9_13.4m}-\ref{fig_polsq_50_166m}. The figures are organized into pairs of lengths: Fig.~\ref{fig_polsq_3.9_13.4m} shows 3.9 and 13.3~m, Fig.~\ref{fig_polsq_20_30m} shows 20 and 30~m and Fig.~\ref{fig_polsq_50_166m} shows 50 and 166~m. For each length the squeezing angle ($\pm0.3^\circ$), squeezing ($\pm0.3$~dB) and antisqueezing ($\pm0.3$~dB) form a column.  Due to the technical limitations of the photodetectors it was not possible to measure above 125~pJ or 20~mW in this particular experimental run.  The losses in this particular set-up were also larger than in that which gave the results in Fig.~\ref{fig:results_13.2m}.

Even though the simulations and experiment agree very well for the angle and the antisqueezing, some small discrepancies appear in the squeezing at longer fibre lengths.  This could be caused by variation of the material parameters along the fibre length, or inaccuracies in the Raman model, which would become more prominent for longer fibres.

Ideal Kerr squeezing should increase with propagation distance.  However the experimental data and simulations show that, above 13.4m, the squeezing at a given input energy is largely insensitive to the length of fibre.  The exception here is that the deterioration of squeezing due Raman effects starts to occur at slightly lower energies.  Thus, the maximum squeezing available at a given fibre length actually decreases for longer lengths.  Meanwhile the antisqueezing increases with propagation distance, as expected.

\subsection{Optimal squeezing as a function of power/length}

\begin{figure}[h!]
\centering
\includegraphics[width = 7cm]{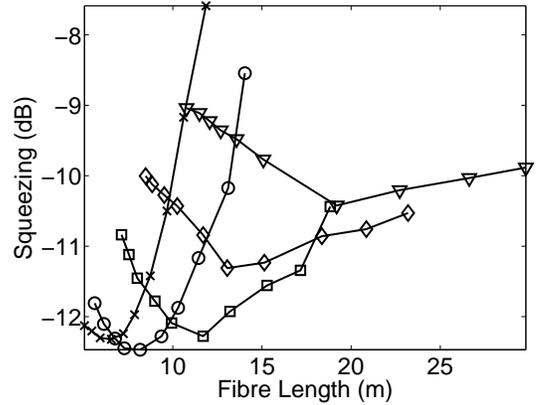}
\caption{Simulated squeezing as a function of fibre length, for various total input energies: $E = $74.4pJ (triangles), $E = $83.7pJ (diamonds), $E = $93pJ (squares), $E = $108pJ (circles) and $E = $119pJ (crosses).  Linear loss and phase-noise have not been included. (Fibre I)}
\label{versuslength}
\end{figure}

Some insight can be gained from further simulations of squeezing as a function of  fibre length, for various input energies, as shown in Fig.~\ref{versuslength}.  This figure reveals that for a given input energy there is an optimum length for the best squeezing, reflecting the length-dependence of the Raman-induced deterioration revealed in the previous plots.  The best squeezing overall is obtained for a pulse at the soliton energy  (54pJ in each pulse), which indicates that the reduced optimal squeezing at other energies is due to pulse-reshaping effects.  Thus for the $t_0=$130fs used here, the optimum fibre length would be $L\simeq$ 7m, (although the improvement over 13m would only be a fraction of a dB).   Alternatively, for a fixed fibre length, one could optimise the maximum squeezing by changing the pulse-width to yield a soliton at that point.

Furthermore, as Fig.~\ref{versuslength} plots the simulation results without adjustment for linear loss, it shows that inferred squeezing of over -12dB is possible.

\section{Conclusion \& Outlook}

An excellent $-6.8\pm0.3$~dB of polarization squeezing, the greatest measured in fibres to date, has been demonstrated with the novel single-pass setup\cite{Dong:2008p116}. From this value it is possible to inferr that $-10.4\pm0.8$~dB of squeezing was generated in the fiber. To further improve the measured noise reduction, losses after the fiber must be minimized by, for example, employing more efficient photodiodes in a minimal detection setup using highest quality optics. We speculate that net losses of as little as 5\% should be possible, thereby allowing the measurement of squeezing in excess of -8~dB. 


By exclusion of the Raman and/or the GAWBS effects in the simulations, it is clear that that the former is a limiting factor for high pulse energies, whereas the latter is detrimental at low energies.  Investigation of a range of fibre lengths revealed that greater squeezing is not achieved going beyond 13.2m.  Indeed, simulations indicate that slightly greater squeezing may be achievable at a lower fibre length of around 7m.


 Further improvement may be possible through the use of photonic crystal fibers (PCF), which are novel fibers manufactured with specially designed light-guiding air-silica structures along their length~\cite{russell03.sci}.  These have already been used in several squeezing experiments~\cite{fiorentino02.ol,lorenz01.apb, hirosawa05.prl,Milanovic:2007p559}, and with fewer low-frequency acoustic vibrations, are also expected to improve squeezing results by minimizing destructive GAWBS noise~\cite{Elser:2006p133901}. Such an advance would bring fiber-produced squeezed states closer to minimum uncertainty states, a desirable feature for quantum information applications.
 

\acknowledgements
The work is funded by the Australian Research Council under the Centres of Excellence scheme, and by the Deutsche Forschungsgemeinschaft.  We gratefully acknowledge useful comments from Murray Olsen and Ben Buchler.

\appendix

\section{Numerical implementation}
\subsection{Absorbing potentials}
The split-step Fourier method used to integrate the equations gives periodic boundary conditions.  Thus any radiation shed by the pulse that reaches the edge of the time window during the simulations will re-enter the other side, eventually interfering with the original pulse and thereby giving possibly spurious results.  To prevent this, we include an inhomogenous loss that absorbs any shed radiation before it reaches the boundary.  We choose a (negative) gain profile of $g(\tau) = \sin^{20}(\pi \tau/2\tau_0)$, where $2\tau_0$ is the width of the simulation window.  The contribution to the Wigner equations is then:
\begin{equation}
\left . \frac{d}{d\zeta} \phi(\tau,\zeta) \right |_{\rm loss} = -g(\tau) \phi(\tau,\zeta) + \Gamma_L(\tau,\zeta), 
\end{equation}
where the loss noise has the correlations
\begin{equation}
\left <  \Gamma_L(\tau,\zeta) \Gamma_L(\tau',\zeta')\right > = \frac{g(\tau)\delta(\zeta-\zeta')\delta(\tau-\tau')}{2\overline n}.
\end{equation}
The $+P$ equations are similar, except that there is no stochastic terms from the loss.

Naturally, because this loss is not a physical effect, the time window should be wide enough so that the loss does not affect the pulse itself.


\subsection{Split Wigner equations}

\label{split_fields}

To increase the output precision in the numerical Wigner calculation, we split the fields into
means plus deviations: $\phi=\overline{\phi}+\delta\phi$, $\beta_{k}=\overline{\beta}_{k}+\delta\beta_{k}$
and evolve the two parts separately. The simulated equations are thus:

\begin{eqnarray}
\frac{d}{d\tau}\overline{\beta}_{k}& = & -ir_{k}^{2}|\overline{\phi}|^{2}e^{i\Omega_{k}\tau}\nonumber\\
\frac{d}{d\tau}\Delta\beta_{k} & = & -ir_{k}^{2}\left\{ \overline{\phi}\Delta\phi^{*}+\Delta\phi\overline{\phi}^{*}+\Delta\phi\Delta\phi^{*}-\frac{vt_{0}}{2\overline{n}z_{0}\Delta\zeta}\right\} e^{i\Omega_{k}\tau}\nonumber\\
\frac{d}{d\zeta}\overline{\phi}& = & \frac{i}{2}\frac{\partial^{2}}{\partial\tau^{2}}\overline{\phi}+i|\overline{\phi}|^{2}\overline{\phi}-iI\overline{\phi}\nonumber\\
\frac{d}{d\zeta}\Delta\phi & = & \frac{i}{2}\frac{\partial^{2}}{\partial\tau^{2}}\Delta\phi+i\left(\overline{\phi}^{*}+\Delta\phi^{*}\right)\left(2*\Delta\phi\overline{\phi}+\Delta\phi^{2}\right) \nonumber \\
 & & +i \Delta\phi^{*}\overline{\phi}^{2} - i\left(I\Delta\phi+\Delta I\overline{\phi}+\Delta I\Delta\phi\right),\nonumber \\
 \end{eqnarray}
 where \begin{eqnarray}
I(\zeta,\tau) & = & \sum_{k}2\Re\left\{ \overline{\beta}_{k}e^{-i\Omega_{k}\tau}\right\} \Delta\Omega\nonumber\\
\Delta I(\zeta,\tau) & = & \sum_{k}2\Re\left\{ \Delta\beta_{k}e^{-i\Omega_{k}\tau}\right\} \Delta\Omega\end{eqnarray}
and with initial conditions \begin{eqnarray}
\overline{\phi}(\zeta=0,\tau) & = & \sqrt{N}\textrm{sech}(\tau)\nonumber\\
\Delta\phi(\zeta=0,\tau) & = & \Gamma_{\phi}(\tau)\nonumber\\
\overline{\beta}_{k}(\zeta,\tau=-\infty) & = & 0\nonumber\\
\Delta\beta_{k}(\zeta,\tau=-\infty) & = & \Gamma_{k}(\zeta).\end{eqnarray}
 The soliton number is defined as $N = E/E_s$, where $E_s = 2\hbar\omega \overline n$ is the energy of a fundamental sech soliton of width $t_0$.

\subsection{Rescaled $+P$ equations}
\label{rescaled_pp}
The rescaled $+P$ equations, corresponding to the Wigner equations of Eq.~(\ref{eq:scaled_wigner}) are
\begin{eqnarray}
\frac{\partial}{\partial\zeta}\phi & = & \lc-i\sum_{k}\left\{ \beta_{k}e^{-i\Omega\tau}+\beta_{k}^{+}e^{i\Omega\tau} \right\} \Delta\Omega +\frac{i}{2}\frac{\partial^{2}}{\partial\tau^{2}}\right.\nonumber\\
&&\left. +i(1-f)\phi^+\phi +\sqrt{i}\Gamma^{E}(\zeta,\tau) +i\Gamma^{R}(\zeta,\tau) \rc\phi\nonumber\\
\frac{\partial}{\partial\tau}\beta_{k} & = & r_{k}^{2}\left(-i\phi^+\phi  +\Gamma^{R}_k(\zeta,\tau) \right)e^{i\Omega\tau},\end{eqnarray}
 with equations of conjugate form for $\phi^+$ and $\beta_k^+$.  The initial conditions are \begin{eqnarray}
\beta_{k}(\zeta,\tau & = & -\infty)=\Gamma_{k}^{\beta}(\zeta)\nonumber\\
\phi(\zeta=0,\tau) & = & \sqrt{\frac{vt_{0}}{\overline{n}}}\ex{\hat{\Psi}(0,t_{0}\tau)},\end{eqnarray}
 where the stochastic terms have correlations 
\begin{eqnarray}
\ex{{\Gamma_{k'}^{\beta}}^{+}(\zeta')\Gamma_{k}^{\beta}(\zeta)} & = & \frac{r_{k}^{2}n_{k}\delta_{k,k'}}{\overline{n}\Delta\Omega}\delta(\zeta-\zeta')\nonumber\\
\ex{\Gamma^{E}(\zeta',\tau')\Gamma^{E}(\zeta,\tau)} & = & \frac{1-f}{\overline n}\delta(\zeta-\zeta')\delta(\tau-\tau')\nonumber\\
&=&\ex{\Gamma^{E+}(\zeta',\tau')\Gamma^{E+}(\zeta,\tau)},\nonumber\\
\ex{\Gamma^{R}(\zeta',\tau')\Gamma^{R}_k(\zeta,\tau)} & = & \frac{1}{\overline n}\delta(\zeta-\zeta')\delta(\tau-\tau')\nonumber\\
&=&\ex{\Gamma^{R+}(\zeta',\tau')\Gamma^{R+}_k(\zeta,\tau)}.\end{eqnarray}
Preliminary investigation of other (physically equivalent) ways to numerically implement the Raman noise did not find any improvement over the simple choice given here.

 
\section{Output Moments}

As discussed in Sec.~\ref{Outputs}, the $+P$ and Wigner simulation methods give correlations that are normally and symmetrically ordered, respectively.  To compare with experimental measurements of the Stokes' parameter variances, some re-ordering is necessary.  The transformation (\ref{rotation}) that relates $\widehat S_1$ to $\widehat S_\theta$ preserves the commutation relations.  Thus to reorder the variance of a Stokes parameter at a general angle in the dark plane, $\widehat{S}_{\theta}^2 $, we only need to consider the corrections that arise from reordering $\widehat{S}_{1}^2$.

\subsection{Normal Ordering}
\label{Normal}

First we expand the mean-square of $\widehat S_1$ as
\begin{eqnarray}
:\widehat{S}_{1}^2: &=&:\widehat{N}_{xx}\widehat{N}_{xx}:-2\widehat{N}_{xx}\widehat{N}_{yy}+:\widehat{N}_{yy}\widehat{N}_{yy}:\;.
\end{eqnarray}
Thus we need only consider the reordering of terms of the form:
 \begin{eqnarray}
\widehat{N}_{\sigma\sigma}\widehat{N}_{\sigma\sigma} &=& \int dz \int dz' \widehat{\Psi}_{\sigma}^{\dagger}(t,z)\widehat{\Psi}_{\sigma}(t,z)\widehat{\Psi}_{\sigma}^{\dagger}(t,z')\widehat{\Psi}_{\sigma}(t,z')\nonumber\\
&=& \int dz \int dz' \widehat{\Psi}_{\sigma}^{\dagger}(t,z)\widehat{\Psi}_{\sigma}^{\dagger}(t,z')\widehat{\Psi}_{\sigma}(t,z)\widehat{\Psi}_{\sigma}(t,z') \nonumber\\
&&+ \int dz \widehat{\Psi}_{\sigma}^{\dagger}(t,z)\widehat{\Psi}_{\sigma}(t,z)
\nonumber\\
&=&:\widehat{N}_{\sigma\sigma}\widehat{N}_{\sigma\sigma}: + \widehat{N}_{\sigma\sigma},
\end{eqnarray}
which gives
\begin{eqnarray}
\widehat{S}_{1}^2 &=&:\widehat{S}_{1}^2: + \widehat{S}_{0}\;.
\end{eqnarray}

\subsection{Symmetric ordering}
\label{Symmetric}

To symmetrically order the relevant products of 4 field operators, we must start from the sum of all $4!=24$ possible orderings.  The essential result that we require is:

 \begin{eqnarray}
\left .\widehat{N}_{\sigma\sigma}\widehat{N}_{\sigma'\sigma'}\right|_{\rm sym} &=& \left(\widehat{N}_{\sigma\sigma}+\frac{1}{2}M\right)\left( \widehat{N}_{\sigma'\sigma'}+\frac{1}{2}M\right)  + \frac{1}{4}M\delta_{\sigma,\sigma'},\nonumber\\
\end{eqnarray}
where $M$ is the number of modes.  The mean square of $\widehat{S}_1$ is then:
\begin{eqnarray}
\left .\widehat{S}_{1}^2\right|_{\rm sym} &=&\left . \left( \widehat{N}_{xx}\widehat{N}_{xx}-2\widehat{N}_{xx}\widehat{N}_{yy}+\widehat{N}_{yy}\widehat{N}_{yy}\right)\right|\;,\nonumber\\
 &=& \widehat{S}_{1}^2 + \frac{1}{2}M.
\end{eqnarray}


\bibliographystyle{ol}
\bibliography{jh_literature,exported_080602,books}

\end{document}